\documentclass[12pt]{article}
\usepackage{url}
\usepackage{subfigure}
\usepackage[labelfont=bf]{caption}
\usepackage{graphicx}
\usepackage{booktabs} 
\usepackage[version=3]{mhchem}
\usepackage{longtable}
\usepackage{amsmath}
\usepackage{breqn}

\usepackage{xr}
\makeatletter
\newcommand*{\addFileDependency}[1]{
  \typeout{(#1)}    
  \@addtofilelist{#1}
  \IfFileExists{#1}{}{\typeout{No file #1.}}
}
\makeatother

\newcommand*{\myexternaldocument}[1]{
    \externaldocument{#1}
    \addFileDependency{#1.tex}
    \addFileDependency{#1.aux}
}
\myexternaldocument{si}

\usepackage{hyperref}
\hypersetup{
    colorlinks=true,
    linkcolor=blue,
    filecolor=magenta,      
    urlcolor=blue,
    citecolor=blue,
    }

\newcounter{lastnote}

\title{The \textit{ab initio} amorphous materials database: Empowering machine learning to decode diffusivity}

\author
{Hui Zheng,$^{1}$ Eric Sivonxay,$^{2, 3}$ Max Gallant,$^{1,2}$ Ziyao Luo,$^{1}$ \\
Matthew McDermott,$^{1, 2}$  Patrick Huck,$^{1}$ Kristin A. Persson$^{1\ast}$\\
\normalsize{$^{1}$Materials Science Division, Lawrence Berkeley National Laboratory, Berkeley, CA, USA}\\
\normalsize{$^{2}$Materials Science and Engineering, University
of California, Berkeley, Berkeley, CA, USA}\\
\normalsize{$^{3}$Energy Technologies Area, Lawrence Berkeley National Laboratory,  Berkeley, CA, USA}\\
\\
\normalsize{$^\ast$E-mail: kapersson@lbl.gov}
}

\date{}

\begin{document} 


\baselineskip24pt


\maketitle 


\begin{abstract}
Amorphous materials exhibit unique properties that make them suitable for various applications in science and technology, ranging from optical and electronic devices and solid-state batteries to protective coatings. However, data-driven exploration and design of amorphous materials is hampered by the absence of a comprehensive database covering a broad chemical space. In this work, we present the largest computed amorphous materials database to date, generated from systematic and accurate \textit{ab initio} molecular dynamics (AIMD) calculations. We also show how the database can be used in simple machine-learning models to connect properties to composition and structure, here specifically targeting ionic conductivity. These models predict the Li-ion diffusivity with speed and accuracy, offering a cost-effective alternative to expensive density functional theory (DFT) calculations. Furthermore, the process of computational quenching amorphous materials provides a unique sampling of out-of-equilibrium structures, energies, and force landscape,  and we anticipate that the corresponding trajectories will inform future work in universal machine learning potentials, impacting design beyond that of non-crystalline materials. 
\end{abstract}

\pagebreak

\section*{Introduction}


Amorphous materials are generally characterized by the lack of long-range order as a result of synthesis processes that freeze in a non-equilibrium, non-crystalline structure. Compared to the stringent synthesis requirements of crystalline materials for ordered atomic arrangement, synthesizing amorphous materials is less energy-intensive as it is often done via rapid quenching or cooling. Furthermore, the rate at which a material is cooled from a liquid to a solid state can influence the properties significantly\cite{debenedettiSupercooledLiquidsGlass2001a}. Such tunability can be leveraged to engineer a wide range of physical, chemical, and mechanical properties. As examples, bulk metallic glasses with unique magnetic properties are found in high-efficiency transformers\cite{Najgebauer2023, KhanRecentAdvancements2018}, amorphous alkali-aluminosilicates were made famous as the world-leading cover glass for portable electronics\cite{gomezLookChemicalStrengthening2011}, and amorphous 2D boron nitrides are proposed for the next-gen memory solutions due to their ultra-low dielectric constant combined with excellent electrical and mechanical properties\cite{hongUltralowdielectricconstantAmorphousBoron2020,khotAmorphousBoronNitride2022}. 

Amorphous materials are also considered for various applications in energy storage. For example, amorphous anodes, particularly silicon and silicon-tin alloys, are pursued as high-capacity, lower-cost alternatives to graphite. \cite{yoonEffectSEILayer2023, majeedSiliconbasedAnodeMaterials2023, zhangChallengesRecentProgress2021, zhangSiliconAnodesImproved2021, hasaElectrochemicalReactivityPassivation2020,beaulieuReactionLiAlloy2003,mcdowell25thAnniversaryArticle2013}  Furthermore, the conformal nature of amorphous materials proffers major advantages in electrode coating applications\cite{chengEvaluationAmorphousOxide2020, chengMaterialsDesignPrinciples2022, sivonxayLithiationProcessLi2020, sivonxayDensityFunctionalTheory2022}  and as electrolytes for all-solid-state batteries.  While crystalline \ce{Li7La3Zr2O12} (LLZO) exhibits high Li-ion conductivity, it also suffers from lithium dendrite growth through the grain boundaries \cite{han_high_2019, grady_emerging_2020, sastre_blocking_2021}. In comparison, amorphous LLZO exhibits lower Li-ion diffusivity but shows marked improvement in safety and cyclability\cite{sastreBlockingLithiumDendrite2021}. In contrast, amorphous lithium phosphorus oxynitride (LiPON) shows a higher Li-ion diffusivity than its crystalline counterpart\cite{sastre_blocking_2021, lacivita_structural_2018, gradyEmergingRoleNoncrystalline2020}, which indicates a possible design space where ionic conductivity and safety can be optimized within an amorphous phase space. 

Unfortunately, measuring ionic diffusivity in inorganic (amorphous or crystalline) materials is highly time-consuming, and more often than not, the inherent bulk diffusivity is masked by other factors, such as pellet densification. It is possible to obtain an estimate of the ionic diffusivity through Ab Initio Molecular Dynamics (AIMD); however, in comparison to crystalline compounds, amorphous materials require larger unit cells to capture sufficient and representative local environments. As of writing, the only reported database is the amorphous nanoporous materials database, which includes 75 amorphous carbons, 119 polymers, and 16 kerogens. The database is curated from earlier literature\cite{thyagarajanDatabasePorousRigid2020} and covers a limited range of compositions. To meet the need to accelerate our discovery and design of amorphous ionic conductors with optimized ionic diffusivity, in this study, we present an extensive, computed amorphous database covering 5120 compositions and 79 elements generated through systematic AIMD calculations. Additionally, we demonstrate the database's applicability in simple yet highly efficient machine-learning models to rapidly and accurately predict Li diffusivity, providing a cost-effective alternative to density functional theory (DFT) calculations. The database also opens new possibilities for improving current directions in universal machine learning potentials by providing unique information about structure-energy-force relationships far from equilibrium configurations.

\section*{Results and Discussion}

\subsection*{Data scope}
\begin{figure}[htp]
\centering\includegraphics[width=1.0\linewidth]{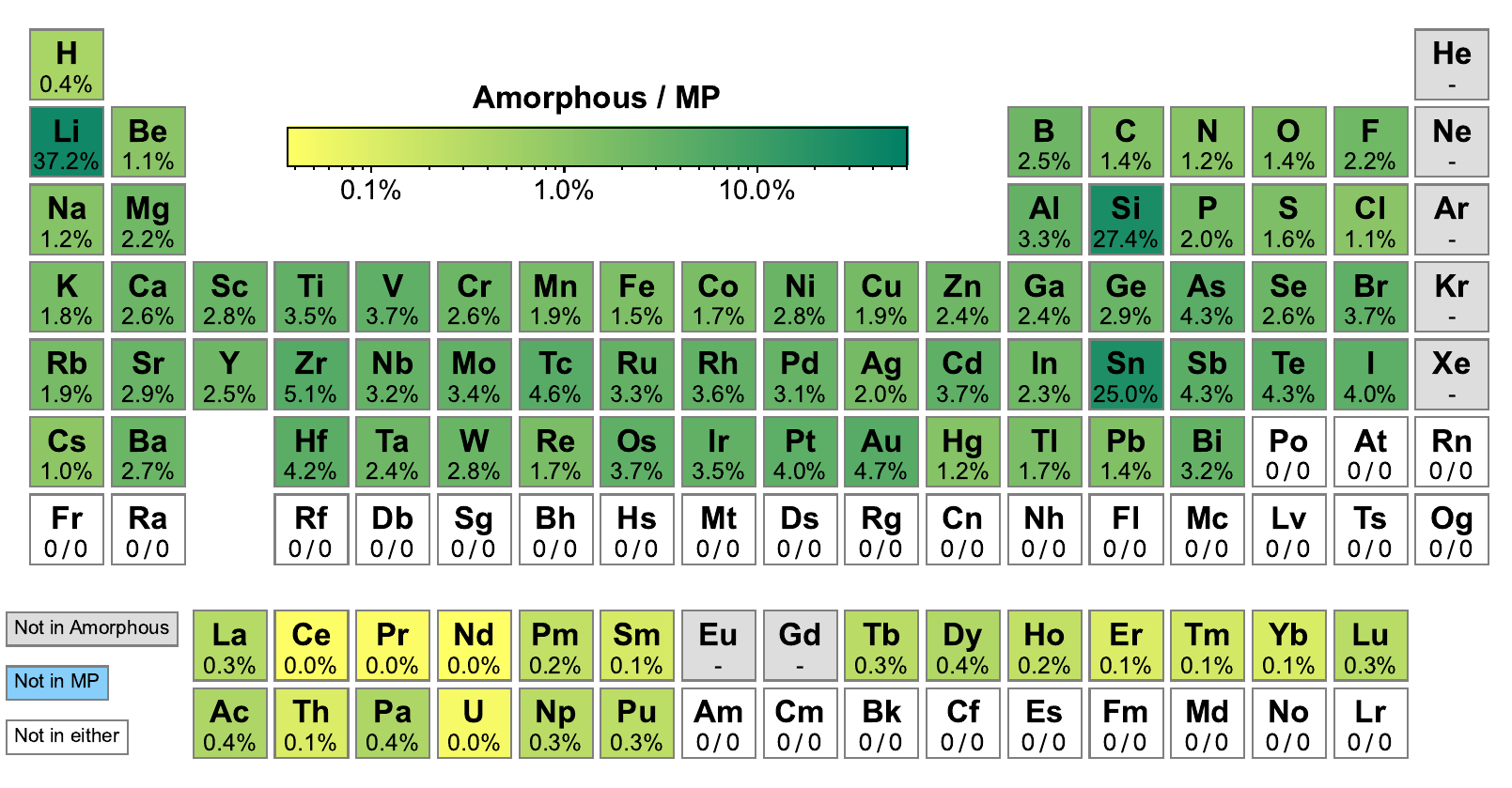}
\caption{\label{fig:data_scope} \textbf{Elemental occurrence in the 5000K amorphous database compared to the Materials Project}. Element occurrence ratios for compositions in the amorphous database are shaded by color scale.}
\end{figure}

The amorphous database includes two subset databases. The first one consists of 5,120 compounds, which are melted at 5000K using the MPMorph workflow. Details about the workflow can be found in \nameref{methods} section. This database is here denoted as the `5000~K amorphous database'. The second database is generated from the last snapshot structures from the 5000K and annealed at 1000K, 1500K, 2000K, and 2500K for selected compositions. This database is denoted as the `multi-temperature amorphous database'. 
Among the 5,120 compositions in the 5000K amorphous database, 3,533 compounds contain lithium (SI Figure \textcolor{blue}{1}). Figure \ref{fig:data_scope} presents the proportion of each element's occurrence within the compositions in the 5000K amorphous database, compared to its occurrence within the Materials Project database. We note that the 5000K database exhibits extensive coverage,  providing a similar representation of compositions compared to the Materials Project. The element occurrence of the compounds in the Materials Project is shown in SI Figure \textcolor{blue}{2}, where there are approximately 50,000 compounds containing Li. Similarly, the element occurrence in the compositions covered in the multi-temperature amorphous database is shown in SI Figure \textcolor{blue}{3}. SI Figure \textcolor{blue}{4} shows the ratio of the element occurrence within the multi-temperature amorphous database compared to its occurrence within the Materials Project database. We find that the multi-temperature amorphous database also effectively captures a diverse range of material compositions, ensuring a comprehensive chemical representation. 

 
\subsection*{Correlations between Li$^{+}$ diffusivity and composition}
Amorphous materials exhibit short-range ordering, which is strongly dependent on the composition. For example, amorphous \ce{Al2O3} exhibits a distribution of 4, 5, and 6-fold oxygen-coordinated \ce{Al^{3+}} units, where the distribution depends on the synthesis or formation conditions. Since cationic diffusion in amorphous materials has been shown to be highly correlated and dependent on bond-formation/breaking events between the cation and its anionic environment\cite{chengEvaluationAmorphousOxide2020, chengMaterialsDesignPrinciples2022}, we anticipate that Li$^{+}$ diffusion in amorphous inorganic materials will correlate strongly to anion specie and composition. In the following section, we identify and analyze the correlations between the Li-ion diffusivity and the i) composition of the materials, ii) size of the anions and cations, and iii) electronegativity difference between the compositional species. 

\begin{figure}[htp]
\centering\includegraphics[width=1.0\linewidth]{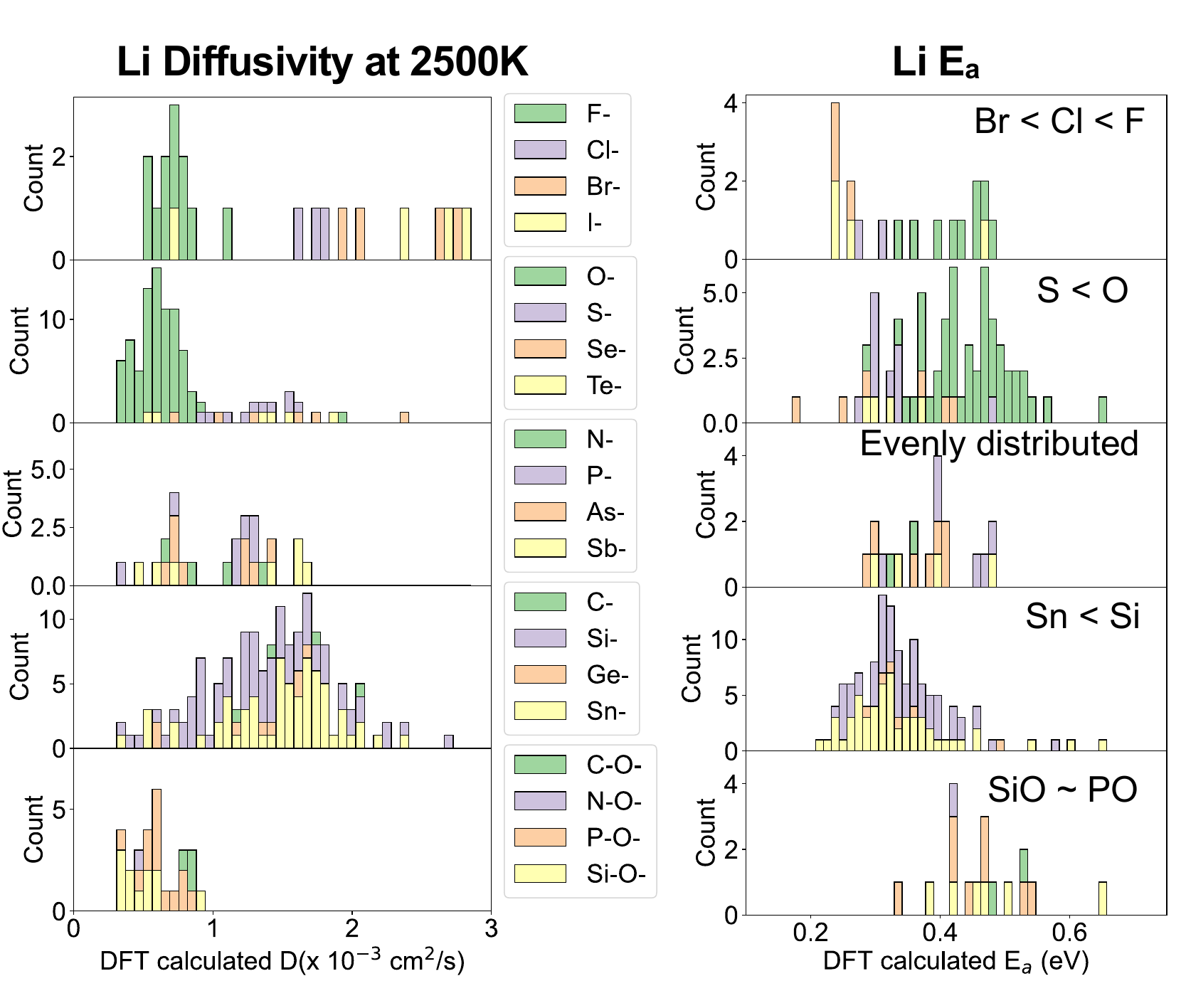}
\caption{\label{fig: Ea_distribution} \textbf{Li diffusivities and activation energies}. Distributions of Li diffusivity, $D$ (2500 K), and activation energy, $E_a$, calculated from the multi-temperature amorphous database. Compositions are sorted based on the anion element present in the system and collated by group on the periodic table. The annotation on the right panel shows the Li $Ea$ order of the peak location from kernel density estimate (KDE) as shown in Supplementary Information (SI) Figure \textcolor{blue}{5}}
\end{figure}

In the context of data coverage, we emphasize that the samples obtained from the collected group may have different sizes. For instance, there is a higher number of compounds containing oxygen compared to those containing sulfur, selenium, and tellurium. Similarly, there is a greater presence of compounds with fluorine compared to compounds containing chlorine, bromine, and iodine. This is demonstrated in Figure \ref{fig: Ea_distribution} and Supplementary Information (SI) Figure \textcolor{blue}{3}. A small sample size may impact the accuracy in comparing different anion groups. Therefore, our analysis focuses on compounds with larger sample sizes, ensuring that the distributions are distinct enough to yield conclusive results.

Figure \ref{fig: Ea_distribution} and SI Figure \textcolor{blue}{5} show the Li diffusivity and activation barrier distributions calculated from the multi-temperature amorphous database. We make several observations of Li ion diffusivity trends within different groups in the Periodic Table.  Within the halogen group, compositions that include fluorine (\ce{F}) demonstrate significantly lower diffusivity and higher activation energy ($E_a$) compared to those containing chlorine (\ce{Cl}), bromine (\ce{Br}), or iodine (\ce{I}). Similarly, in the chalcogen group, compositions incorporating oxygen (\ce{O}) exhibit lower diffusivity and a higher $E_a$ when compared to those that include sulfur (\ce{S}). Compounds containing elements from group VA group have evenly distributed $E_a$ due to the small sample size. Compounds with tin (\ce{Sn}) have slightly lower $E_a$ compared to those with silicon (\ce{Si}). These trends all follow a similar pattern such that a larger atomic radius of the anion species -- corresponding to elements from a larger row number within the same group -- results in lower Li activation energy ($E_a$) and, thus, higher Li diffusivity. Correspondingly, higher electronegativity or higher charge density leads to stronger bonding between Li and anions, resulting in higher $E_a$ for Li diffusion. The even distribution of activation energies ($E_a$) among compounds containing oxyanions, as shown in the bottom distribution, may be attributable to the small sample size.

In addition to the correlation between anion species and Li diffusivity, the presence of other cations can also influence Li diffusivity. SI Figure \textcolor{blue}{6} depicts the average Li diffusivity values from the 5000~K amorphous database across compounds containing specific elements from the periodic table. Certain elements correspond to higher Li diffusivity than others. For example, there are two regions of elements that contribute to high Li diffusivity: the alkali/alkaline metals group (IA and IIA) and the right-hand side of the periodic table, encompassing groups IB, IIB, and IIIA through VIIA. For compounds containing elements from these groups, a trend is observable: with an increasing row number (and hence, larger atomic radius), Li diffusivity also increases. The presence of cations originating from groups IIIA to VIIA will likely result in polyanionic environments (carbonates, nitrates, phosphates,polyhalogens, etc.) within the amorphous material, which on average leaves the Li$^{+}$ less directly coordinated to oxygen and hence more free to move. SI Figure \textcolor{blue}{7} further illustrates the standard deviation (STD) of Li diffusivity for compounds containing specific periodic elements. Notably, compounds that incorporate elements from groups IIB and IIIA to VIA demonstrate smaller STDs when compared to compounds containing alkali elements.

SI Figures \textcolor{blue}{8} and \textcolor{blue}{9} present similar plots for the activation barrier ($E_a$) of Li$^{+}$, derived from the more limited multi-temperature amorphous dataset. Owing to the smaller sample size of the data, the distribution of elements associated with lower Li $E_a$ is not as pronounced as the Li diffusivity distribution from the 5000K database displayed in SI Figure \textcolor{blue}{6}. Nevertheless, compounds containing Cl, Br, I, S, Se, Pb, Sr, Sn, In, Ba, Na, K, and Rb demonstrate lower $E_a$ than other compounds. This observation aligns with cases of high Li diffusivity in the 5000K database, as depicted in SI Figure \textcolor{blue}{6}.

\begin{figure}[htp]
\centering\includegraphics[width=1.\linewidth]{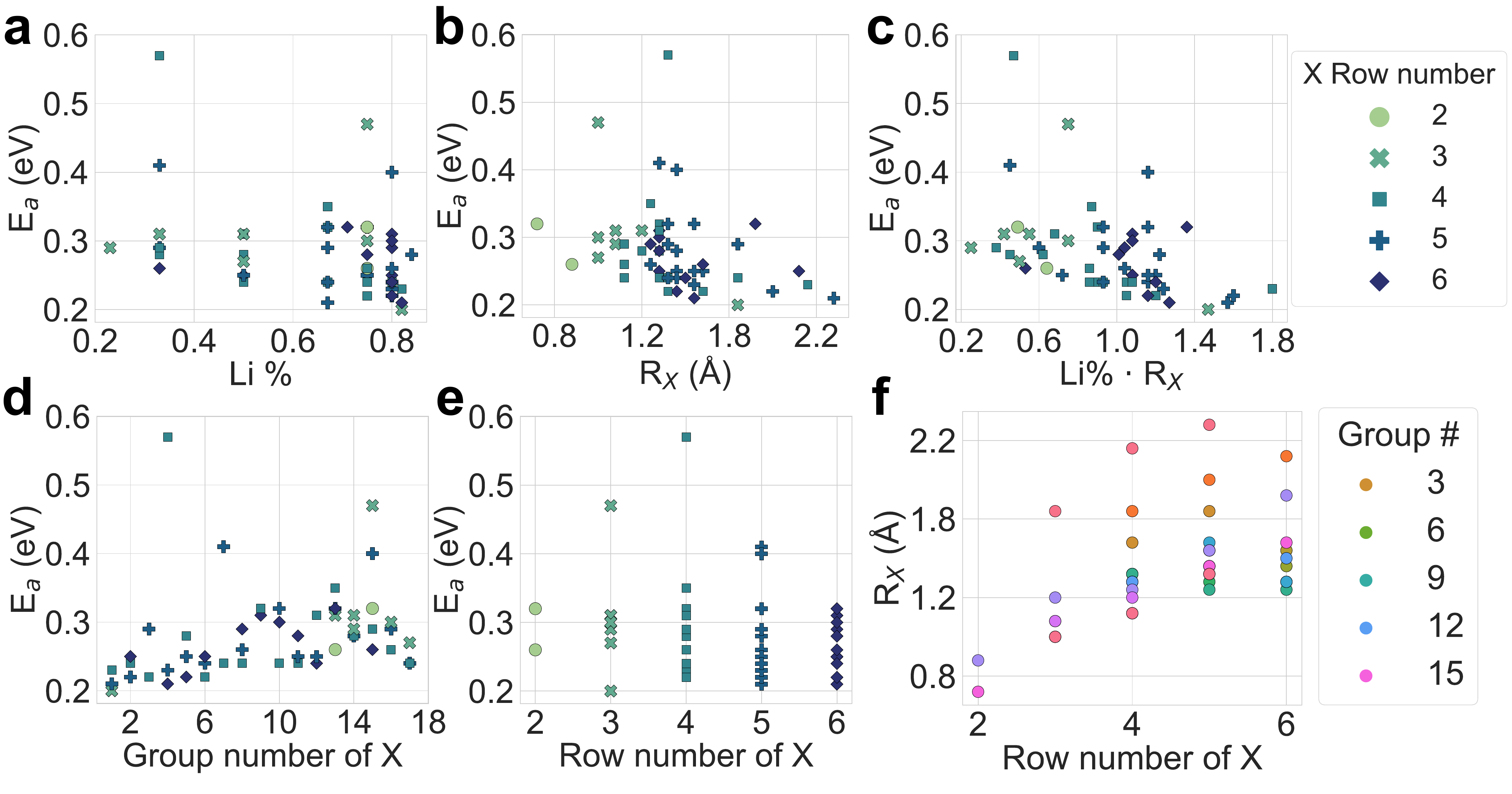}
\caption{\label{fig: binary}\textbf{Effect of composition on Li activation energies in binary compounds}. \textbf{a} - \textbf{e} show the correlations between the activation energy $E_a$ of Li in binary compounds and the elemental properties of the coexisting species $X$ and Lithium fraction (Li\%) in the binary composition (\ce{LiX}), such as atomic radius of element $X$ ($R_X$), the multiplication of these two (Li\%$\cdot R_X$), group number of $X$, row number of $X$. Color and marker shapes signify the row number of element $X$. \textbf{f} plots the correlation between the $R_X$ and row number of $X$; colors are used to distinguish group numbers of element $X$.}
\end{figure}

Our analysis encompasses compounds that range from binary to ternary, quaternary, and even quinary. Consequently, Li-ion diffusivity associated with one element often cross-correlates with other elements present in the same set of compounds. This necessarily results in some over-counting and cooperative effects on Li diffusivity or $E_a$. To deconvolute these relationships, we focus on available binary \ce{Li$X$} compounds (where $X$ represents any species within the composition). This approach allows us to clarify the correlations between Li diffusivity and other elements. Figure \ref{fig: binary} illustrates the correlation between the activation barrier of Li ($E_a$) and the properties of the $X$ species in \ce{Li$X$} compounds. Panel \textbf{a} reveals a negative correlation between $E_a$ and the Li fraction in the composition. This observation corresponds to a similar phenomenon found in crystalline solid-state electrolyte systems, denoted 'Li stuffing', where increased Li content improves Li diffusion\cite{xiaoLithiumOxideSuperionic2021}. Panel \textbf{b} demonstrates a strong negative relationship between Li $E_a$ and the atomic radius of $X$ ($R_X$), showing that larger $X$ species facilitates Li diffusion by providing more spacious frameworks and - in the case of anions, lowers the electronegativity. In SI Figure \textcolor{blue}{10}, the color gradient further clarifies why some large-radius points do not exhibit correspondingly small $E_a$: a lower Li percentage. Thus, for compounds with similar Li percentages, a larger atomic radius of $X$ implies lower $E_a$. Figure \ref{fig: binary} panel \textbf{c} presents an approximately linear negative correlation between Li $E_a$ and the product of Li fraction (Li\%) and atomic radius of $X$ $R_X$. Panels \textbf{d} and \textbf{e} explore the influence of $X$ species' group and row numbers on Li $E_a$ in \ce{LiX}. As the group number of $X$ and the corresponding electronegativity difference between Li and $X$ increase, Li $E_a$ tends to rise due to stronger Li-$X$ bonds, thus inhibiting Li diffusion. Although an increase in $X$ row number generally results in lower mean values of Li $E_a$, the impact is not pronounced due to the wide $E_a$ distribution within species of the same row number. Panel \textbf{f} highlights the trend that larger $X$ row numbers correlate with larger atomic radii, with the hue distinguishing $X$ species from different groups.

\subsection*{Feature design for machine learning}
As previously analyzed, both compositional and elemental properties of species correlate with Li diffusivity. However, other features also directly or indirectly impact Li diffusivity. Building on the work of Sendek et al. \cite{sendekHolisticComputationalStructure2017}, who developed a feature set to differentiate high- and low-diffusivity materials, we have expanded the feature set to include more compositional features. This approach equips the machine learning models to learn the underlying correlations between the features and Li diffusivity more comprehensively. The expanded list of features, sorted by their Pearson correlation coefficient, is provided in Table \ref{table: feature}. The compositional features added to the feature set here include Li percentage (Li\%), which has a significant positive correlation with the Li diffusivity in different compositions. A similar trend has been observed in amorphous coating materials for Li-ion batteries as well as crystalline solid-state electrolyte materials\cite{chengEvaluationAmorphousOxide2020, chengMaterialsDesignPrinciples2022, xiaoLithiumOxideSuperionic2021}. 
The weighted average of cohesive energy $\overline{E_{coh}}$ is calculated from the cohesive energy of the ground state of the constituent elemental systems. The cohesive energy provides a useful metric for describing the average bond strength of the local units in the amorphous material. Specifically, the stronger the bonds, the harder it is for the activated bond-breaking process to occur, which underpins the diffusion process.  Hence, $E_{coh}$ negatively correlates with the Li diffusivity. In addition, a few other features show a negative correlation with Li diffusivity and can be explained in terms of the packing fraction of a set of non-lithium (non-Li) atoms that are in close proximity to a central lithium (Li) atom within a specified radius. We here denote these features as i) set-of-non-Li-atoms packing fraction (SPF), ii) set-of-non-Li-atoms neighbor count (SNC), iii) Li neighbor count (LNC), iv) Li bond ionicity (LBI), and v) density. These structural features also show a strong negative correlation with Li diffusivity, as when the non-Li atoms are parking closely and form a tight structural motif framework, it is harder for Li to diffuse. For more details on the quantitative definition of other features, such as weighted average bulk moduli ($\overline{B}$) and electronegativity ($\overline{X}$), please refer to table \ref{table: feature}, supplementary materials, and SI of reference by Sendek et al.\cite{sendekHolisticComputationalStructure2017}.

\begin{longtable}[c]{llrl}
\caption{Pearson correlation coefficients between various features and the Li diffusivity, obtained from 5000K AIMD calculations. The coefficients are categorized into two bins—Positive and Negative—and the values within each bin are sorted by the Pearson correlation coefficients.\label{table: feature}}\\
\toprule
Feature & Feature Description  &   Pearson $r$ &   Unit  \\
\midrule
$\overline{E_{coh}}$ & weighted cohesive energy               &      -0.81 &     eV \\
SPF & set-of-non-Li-atoms packing fraction  &      -0.76 &      1 \\
density &   weight/volume                &      -0.67 &  g$\cdot$cm$^{-3}$  \\
LNC & Li neighbor count            &      -0.64 &      1 \\
$\overline{B}$ & weighted bulk modulus                 &      -0.64 &    GPa \\
$\overline{X}$ & weighted electronegativity &      -0.64 &      1 \\
SNC & set-of-non-Li-atoms neighbor count   &      -0.63 &      1 \\
LBI & Li bond ionicity                       &      -0.61 &      1 \\
AFC & anion framework coordination                              &      -0.54 &      1 \\
$\overline{E_{coh}^{Li}}$ & weighted cohesive energy exclude Li               &      -0.53 &     eV \\
SLPE & straight-line path electronegativity                            &      -0.51 &      1 \\
PF & packing fraction                               &      -0.46 &      1 \\
SDLI & standard deviation of Li bond ionicity          &      -0.34 &      1 \\
$\overline{m}$ & weighted atomic mass                    &      -0.32 &     kg \\
$\overline{G}$ & weighted shear modulus   &      -0.29 &    GPa \\
ENS & electro-negativity of set-of-non-Li-atoms                            &      -0.29 &      1 \\
$\overline{X^{Li}}$ & weighted electronegativity exclude Li  &      -0.29 &    GPa \\
$\overline{B^{Li}}$ & weighted bulk modulus exclude Li    &      -0.28 &    GPa \\
LLSD & Li-Li separation distance        &      -0.27 &      Å \\
RBI & ratio of LBI and SBI                            &      -0.24 &      1 \\
$\overline{G^{Li}}$  & weighted shear modulus exclude Li       &      -0.10 &    GPa \\
SDLC & standard deviation in Li neighbor count   &      -0.08 &      1 \\
$\frac{\overline{R^{Li}}}{\Bar{R}}$   &  the ratio of the average radius without Li and with Li      &      -0.04 &      1 \\
$\Bar{R}^2\cdot\sqrt{\frac{\Bar{B}\cdot\Bar{R}}{\Bar{m}}}$-Li & synthetic feature (exclude Li)            &      -0.01 &   m$^2$/s \\\hline
\
Li\% & Li percentage                       &       0.77 &      1 \\
SLPW\_pp  & average straight-line path width (point-to-point)  &       0.59 &      Å \\
AAV & average atomic volume        &       0.58 &     Å$^3$ \\
$\overline{R}$  & weighted average atomic radius                &       0.56 &      Å \\
$\overline{X_{Li} - X_{others}}$  &  electronegativity difference  &       0.54 &      1 \\
LLB & Li-Li bonds per Li     &       0.49 &      1 \\
LASD & Li-anion separation distance &       0.49 &      Å \\
SLPW & average straight-line path width    &       0.48 &      Å \\
VPA & volume per anion  &       0.44 &    Å$^3$ \\
AASD & minimum anion-anion separation distance &       0.36 &      Å \\
$\overline{R^{Li}}$  & weighted average atomic radius (exclude Li)     &       0.35 &      Å \\
$\sqrt{\frac{E_{coh}\cdot\Bar{R}^2}{\Bar{m}}}$  & synthetic feature                    &       0.29 &   m$^2$/s \\
SBI & set-of-non-Li-atoms bond ionicity  &       0.26 &      1 \\
$\overline{m^{Li}}$  & average mass  (exclude Li)    &       0.21 &     kg \\
$\Bar{R}^2\cdot\sqrt{\frac{\Bar{B}\cdot\Bar{R}}{\Bar{m}}}$ & synthetic feature  &       0.18 &  m$^2$/s \\
 $\Bar{R}^2\cdot\sqrt{\frac{\Bar{G}\cdot\Bar{R}}{\Bar{m}}}$  & synthetic feature &       0.06 &  m$^2$/s \\
RNC & ratio of LNC and SNC &       0.01 &      1 \\
 $\Bar{R}^2\cdot\sqrt{\frac{\Bar{G}\cdot\Bar{R}}{\Bar{m}}}$-Li  & synthetic feature (exclude Li)   &       0.01 &  m$^2$/s \\
\bottomrule
\end{longtable}

\subsection*{Machine learning models}
\begin{figure}[htp]
\centering\includegraphics[width=0.95\linewidth]{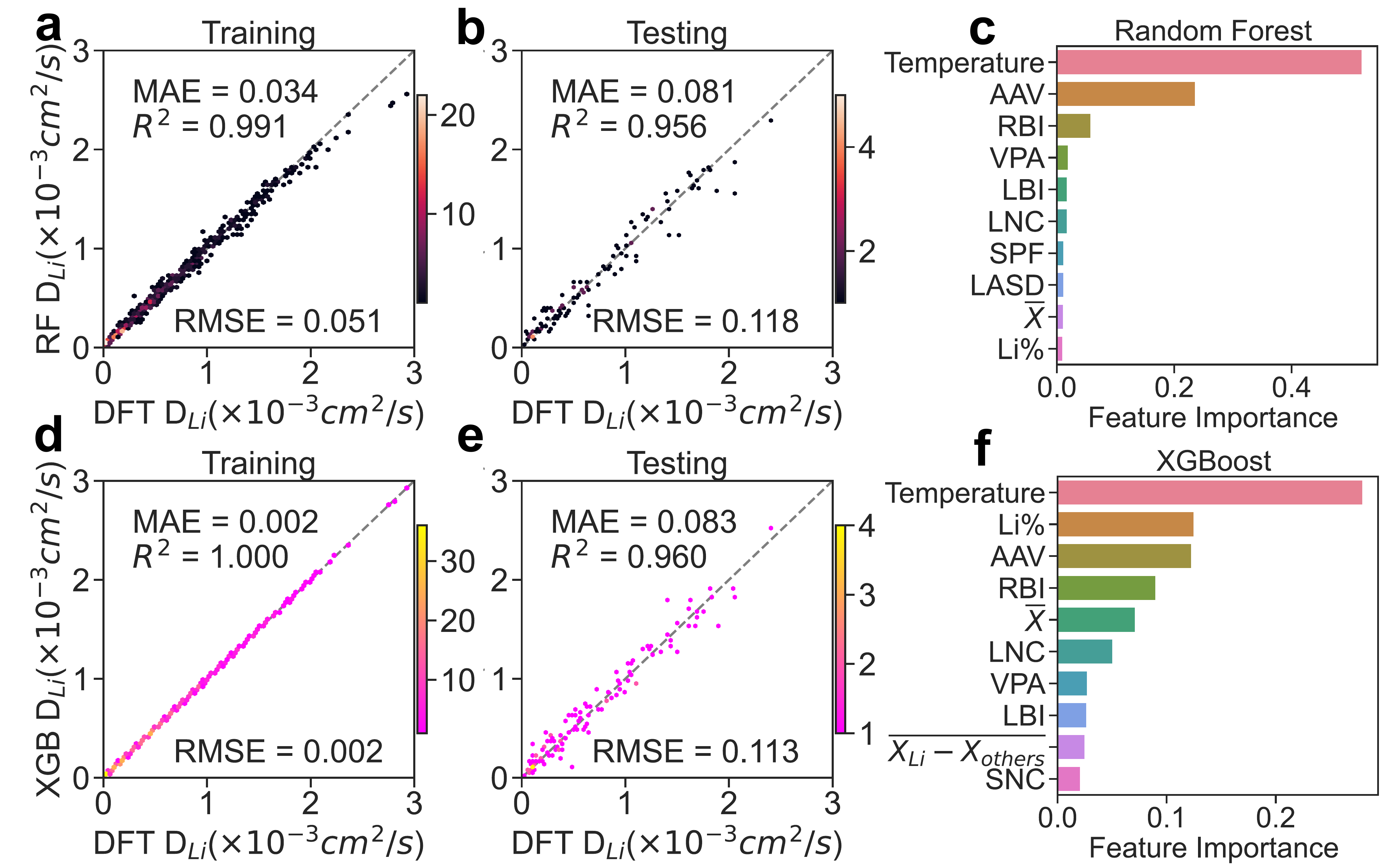}
\caption{\label{fig: ml models} \textbf{Performance of ensemble learning models in diffusivity prediction}. The random forest and XGBoost models are used to predict the Li diffusivity. Parity plots between the DFT calculated Li diffusivity, and the random forest model (\textbf{a} for training and \textbf{b} for testing) and XGBoost model (\textbf{d} for training and \textbf{e} for testing) predicted Li diffusivity, respectively. \textbf{c} and \textbf{f} ranked the top 10 important features analyzed from the random forest and XGBoost models, respectively.} 
\end{figure}

Three different diffusivity-prediction models are trained on the amorphous database. Two ensemble learning models, Random Forest (RF)\cite{ho1995random} and Extreme Gradient Boosting (XGBoost) \cite{ChenXGBoost2016}, were employed to learn the temperature-dependent diffusivity. The parity plots comparing the DFT calculated Li diffusivity and ML-predicted Li diffusivity for both training and test data are shown in Figure \ref{fig: ml models}. We observe that both algorithms achieve coefficients of determination ($R^2$) close to 1, very low mean absolute error (MAE), and Root Mean Squared Error (RMSE), as annotated on the plots. Five-fold cross-validations have been used to assess the performance and the generalization ability of these two models via Scikit-learn\cite{sklearn_api}. Both RF and XGBoost measure feature importance, which indicates the relative significance of a particular feature in diffusivity prediction. The top 10 most relevant features identified from RF and XGBoost models are shown in Figure \ref{fig: ml models} \textbf{c} and \textbf{f}, respectively.  Somewhat trivially, both models rank temperature as the most important feature for predicting the diffusivity at different temperatures. Further, while the orders of the important features predicted from RF and XGBoost may differ, both models share similar highly ranked features: average atomic volume (AAV), Li percentage in the compositions (Li\%), the ratio (RBI) of average Li bond ionicity (LBI) with average bond ionicity of set-of-non-Li-atoms (SBI),  average Li neighbor count (LNC), etc. The definitions of features can be referenced to Table \ref{table: feature}, SI, and Ref~\cite{sendekHolisticComputationalStructure2017} for details. 

\begin{figure}[htp]
\centering\includegraphics[width=1.12\linewidth]{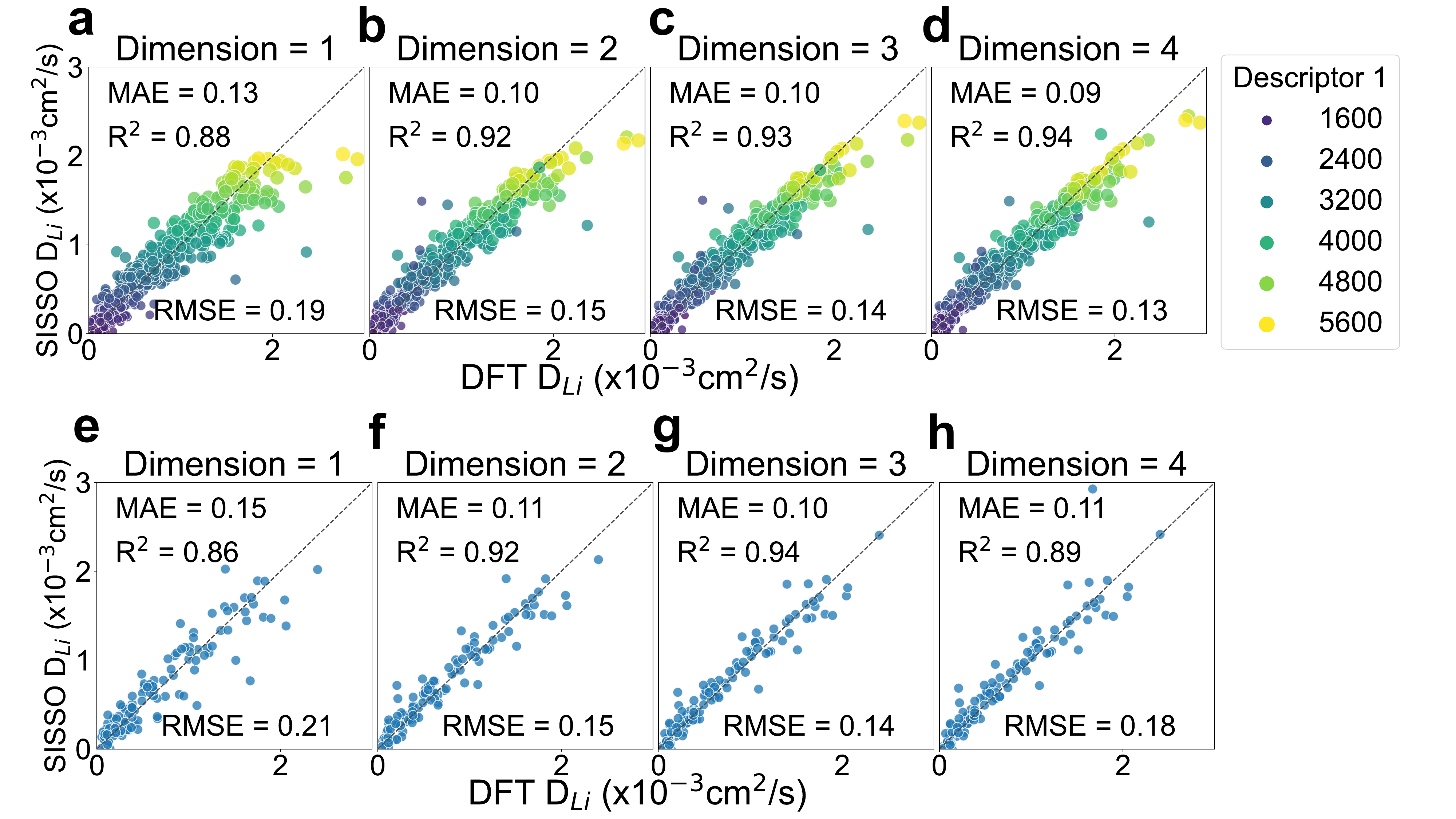}
\caption{\label{fig: sisso models} \textbf{Performance of SISSO models with different complexities in diffusivity prediction.} Parity plots between the DFT calculated Li diffusivity and the SISSO model predicted Li diffusivity. \textbf{a-d} correspond to the parity plots for training data. The colors indicate the values of the first descriptor in the SISSO model, i.e., the first term in Equation (\ref{eq1}). Subfigures \textbf{e-h} are the parity plot for the test data. The complexity of the SISSO model increases from  \textbf{a}-\textbf{d}. I.e., the dimension of the descriptor (a hyperparameter) increases from one to four. $n$-dimensional descriptor means the set of features selected by the $n$ nonzero components of the solution vector \textbf{c}.\cite{ouyangSISSOCompressedsensingMethod2018} }
\end{figure}

Finally, we developed a descriptor for Li$^{+}$ diffusivity using the sure independence screening and sparsifying operator (SISSO) method \cite{ouyangSISSOCompressedsensingMethod2018}. The SISSO model training and prediction results are shown in \ref{fig: sisso models}. The model is capable of capturing the relationship between features and Li diffusivity, even at the simplest model level, with one-dimensional descriptors for training data, as shown in Figure \ref{fig: sisso models} \textbf{a}. However, for the test dataset, when the dimension increases from 1 to 4, the RMSE does not decrease monotonically; instead, it reaches the minimum when the dimension is 3, indicating overfitting beyond this point. As a result, we finalized our SISSO model at the 3-dimensional level. The analytical equation for the three-dimensional model is as follows:
 
\begin{dmath}\label{eq1}
\widehat{D_{Li}} =  4.4\times 10^{-4} \cdot
 T \cdot \frac{SPF + Li\%}{PF}  - 2.1\times \frac{\frac{\overline{R}^2\cdot\sqrt{\frac{\overline{B}\cdot\overline{R}}{\overline{m}}}-Li}{LLB}}{LLB-LNC} -0.28 \times (Li\%-SBI)\times (SDLI-\frac{\overline{R^{Li}}}{\overline{R}}) -0.67
\end{dmath}

The first term of this equation has demonstrated a satisfactory prediction for Li diffusivity ($D_{Li}$), with positive correlations to temperature ($T$) and Li percentage (Li\%) and a negative correlation to the structure's packing fraction (PF). This aligns with the Pearson correlation coefficients sign, indicating that a higher Li percentage, elevated temperature, and looser packing fraction tend to augment Li diffusivity. The SPF and PF ratio also shows a positive correlation with $D_{Li}$. The equation's second term reveals a negative correlation between $D_{Li}$ and an artificially designed synthetic term ($\Bar{R}^2\cdot\sqrt{\frac{\Bar{B}\cdot\Bar{R}}{\Bar{m}}}-Li$) with the same unit as diffusivity. In this term, $\overline{R}$, $\overline{B}$, and $\overline{m}$ symbolize the weighted average of the atomic radius, bulk modulus, and atomic mass, respectively, for the non-Li-atoms composition. As such, it logically follows that for the neighboring atoms of Li, a larger atomic radius and higher bulk modulus of their corresponding elements make the diffusion of Li more challenging. This negative correlation also depends on the difference between Li-Li bonds and Li-neighbor counts (LLB-LNC). While the second and third terms marginally improve the prediction for $D_{Li}$, the first term primarily determines the overall trend.
 

\subsection*{Application of universal machine learning potentials}
\begin{figure}[htp]
\centering\includegraphics[width=0.95\linewidth]{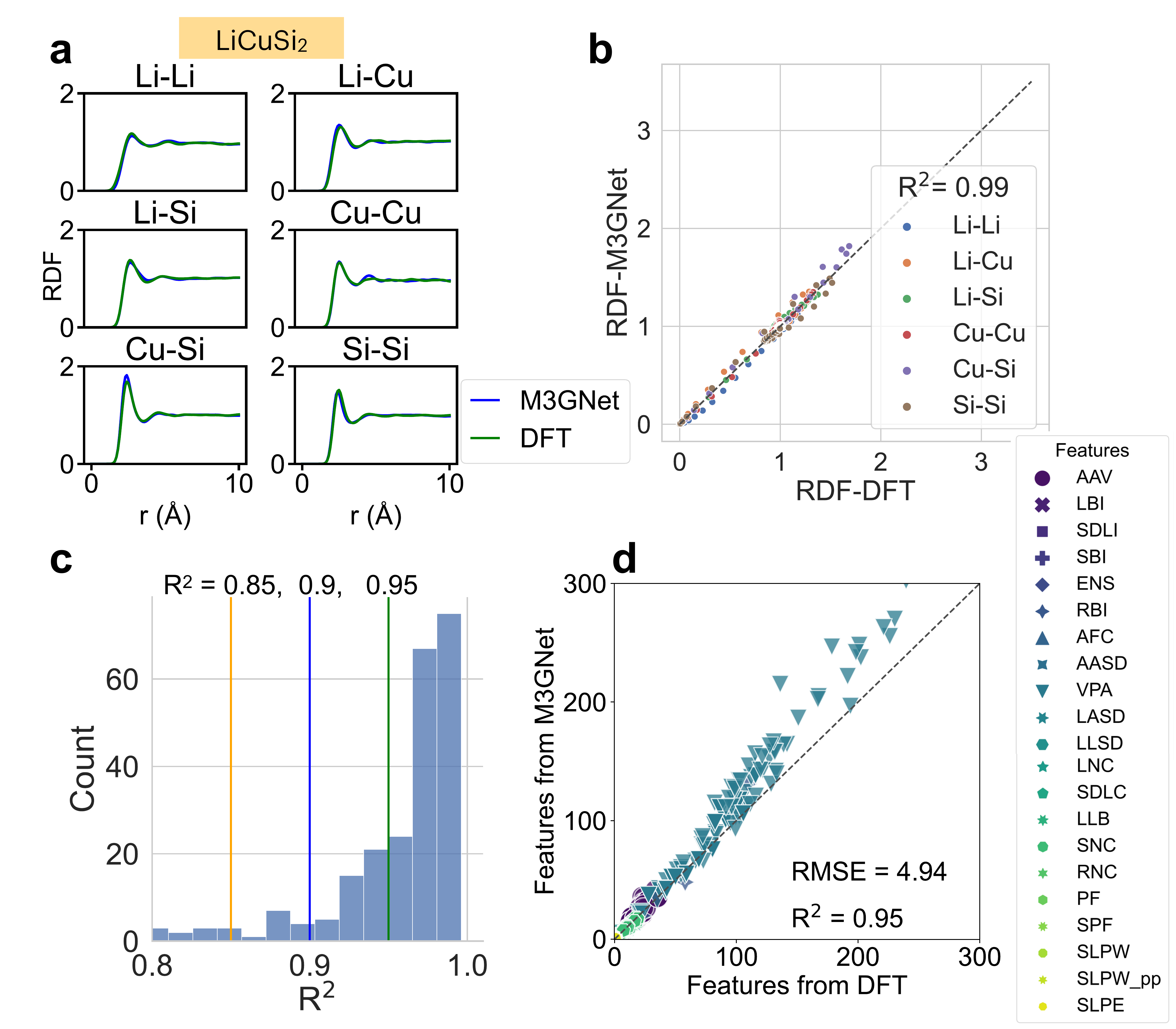}
\caption{\label{fig: RDF}\textbf{The performance of M3GNet in reproducing the amorphous structures from 5000K MD.} \textbf{a} pair-wise radial distribution function (RDF) of \ce{LiCuSi2}, \textbf{b} parity plot comparing DFT-calculated RDF against M3GNet-calculated RDF, with the coefficient of determination regression score ($R^2$) annotated in the legend, \textbf{c} distribution of $R^2$ across the 245 samples of the composition, \textbf{d} parity plot for structure features calculated from AIMD and M3GNet at 5000K.}
\end{figure}

Here, we explore whether the universal interatomic potential M3GNet \cite{chenUniversalGraphDeep2022a}, and CHGNet \cite{dengCHGNetPretrainedUniversal2023} can be used as the surrogate for AIMD calculations to generate the amorphous structures of any composition at a specific temperature. Figure \ref{fig: RDF} \textbf{a} shows the pair-wise RDF comparison between AIMD and M3GNet calculated structures for \ce{LiCuSi2} as an example; the parity plot is shown in Figure \ref{fig: RDF} \textbf{b}, with the $R^2$ equals to 0.99, very close to 1. For 245 samples of amorphous compositions, the distribution of $R^2$ of RDF comparison is plotted in Figure \ref{fig: RDF} \textbf{c}, where 91\% of samples exhibit a $R^2 >$ 0.85, 85\% of samples show $R^2 > $ 0.9, and 68\% of samples manifest $R^2 >$ 0.95. The parity plot of structure feature comparison is shown in Figure \ref{fig: RDF} \textbf{d} with a decent $R^2$ of 0.95. The scales of the structure features have a wide range; the parity plots with different scales can be found in SI Figure \textcolor{blue}{11}. At all ranges, the M3GNet MD calculations can reproduce the AIMD-calculated structures. Therefore, we find that M3GNet is able to generate reasonable amorphous structures and calculate structure features as inputs for the SISSO model to predict Li diffusivity. However, while M3GNet is able to reproduce the Li diffusivity as shown in SI Figure \textcolor{blue}{12}\textbf{a} decently well for the temperature range from 1000K to 2500K, it fails to reproduce AIMD-calculated Li diffusivity at high temperatures (5000K). Therefore, it is suggested that M3GNet be used as a surrogate for AIMD calculations to generate amorphous structures and then used to calculate structure features for the SISSO model to predict Li diffusivity. By employing M3GNet-based molecular dynamics (MD), the calculations achieve a significant speedup of approximately 2000 times (in CPU hours) compared to traditional AIMD methods for diffusivity calculation. 


\section*{Conclusions}

We have developed a comprehensive database for amorphous materials, employing precise but computationally intensive AIMD calculations for over 5120 compositions, spanning from binary alloys \ce{Na2Li9}, \ce{RbLi2}, \ce{SrLi4}, \ce{Li4Zr}, and  \ce{Li4Ta} to ternary, quaternary compounds like \ce{Li4(SiI3)3}, \ce{Sr2Li12Sn}, \ce{Li5La3Nb2O12} and \ce{Li20Si2NiSn2}. This database provides a robust platform for various streamlined machine-learning models, enabling rapid and accurate predictions of ionic diffusivity, here demonstrated for Li$^{+}$ and relevant for applications such as protective coatings and solid-state electrolytes. Universal potentials such as M3GNet and CHGNet, which are predominantly trained on crystalline relaxation trajectories, significantly accelerate calculations compared to traditional AIMD methods and are found to perform well in structure generation but less so for providing ionic diffusivity data. The publication of this database provides unique information about structure-energy-force relationships far away from equilibrium configurations, and we anticipate that it will be a valuable asset in the pursuit of superior universal potentials applicable to non-crystalline materials.

\section*{Methods} \label{methods}

\subsection*{DFT workflow}
\begin{figure}[htp!]
\centering\includegraphics[width=1.\linewidth]{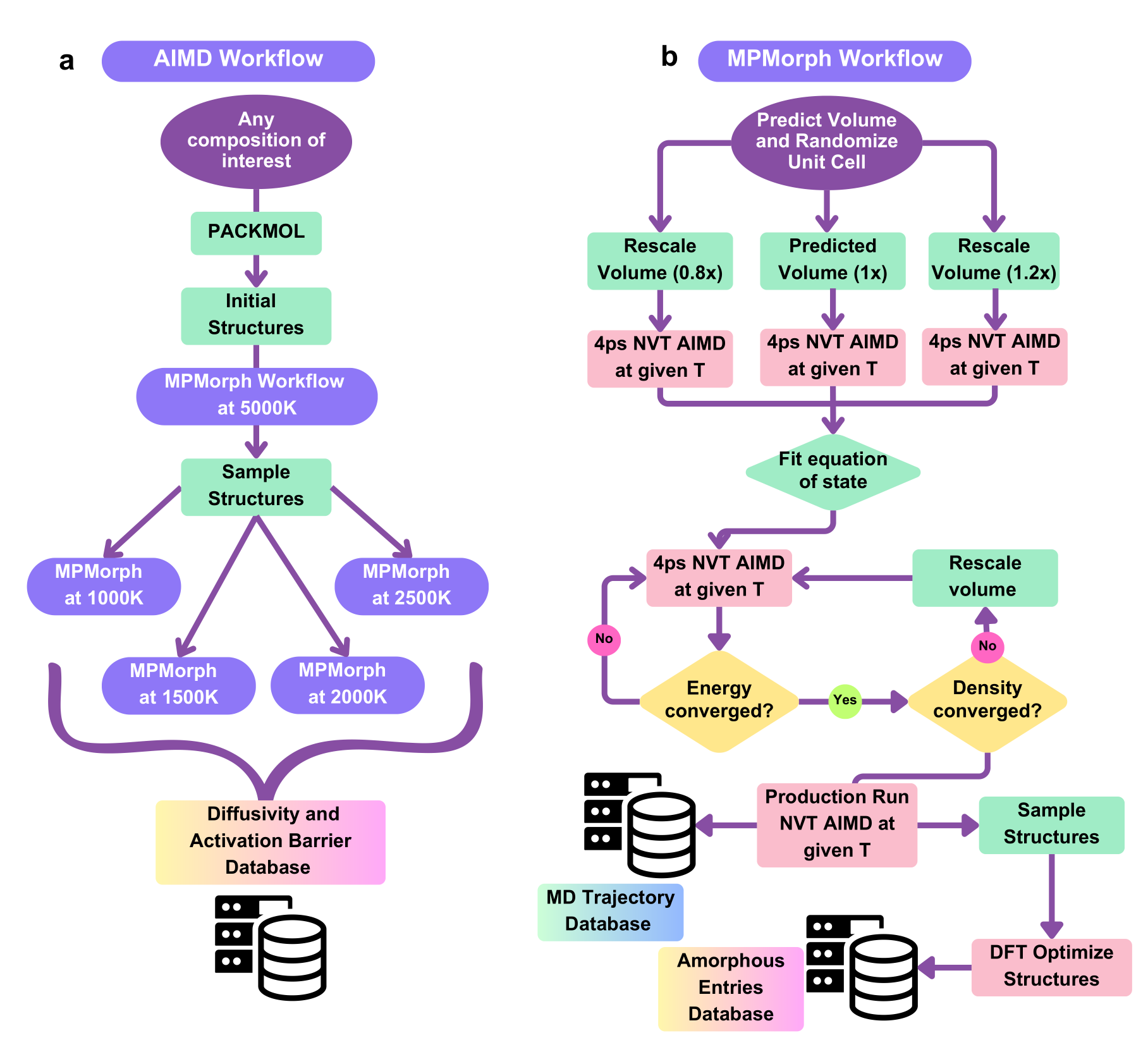}
\caption{\label{fig: workflow} \textbf{Amorphous database workflows}. \textbf{a} Overview of the process used to generate the 5000K amorphous database and the multi-temperature amorphous database. \textbf{b} The MPMorph workflow procedure, which involves identifying the equilibrium volume using the NVT ensemble prior to executing the production run.}
\end{figure}

The database is generated through a combination of well-benchmarked \cite{aykolThermodynamicLimitSynthesis2018} AIMD and MPMorph workflows (see Figures \ref{fig: workflow} \textbf{a}, \textbf{b}), which are designed to generate series of samples of amorphous structures and their respective dynamic behavior at a range of temperatures. The structure sample generation uses PACKMOL\cite{martinezACKMOLPackageBuilding2009} to approximate an initial random structure for a given composition of interest. Subsequently, the MPMorph workflow (Figures \ref{fig: workflow}\textbf{b} ) scales the volumes to 0.8 and 1.2 times the initial volume, performing a 4 ps NVT AIMD run to fit the equation of state at the specified temperature.
A tentative volume is then used to execute another 4 ps NVT AIMD run, ensuring the energy and density have converged. If convergence is achieved, a 20 ps AIMD ``production'' run is conducted using this volume. If not, the workflow iteratively rescales the volume until a value that ensures energy and density convergence is identified. This converged volume is then employed for the 20 ps production run. As shown in Figures \ref{fig: workflow}\textbf{a}, the 5000~K NVT runs have so far generated amorphous structures for more than 5000 compositions. The database corresponding to the 20 ps 5000K NVT run trajectories is denoted the ``5000~K amorphous database''. The last snapshot structure from the 5000K run is used as the input structure for MPMorph workflow at 1000K, 1500K, 2000K, and 2500K to generate the multi-temperature amorphous database. 

The Amorphous Diffusivity Database is made accessible to the public via the Material Project's MPContribs\cite{huckUserApplicationsDriven2016} website \url{https://contribs.materialsproject.org/projects/amorphous_diffusivity} and advanced application programming interface (API) with a dedicated Python client\cite{mpcontribs-client}. Ten structures are sampled every 2 ps for DFT relaxation and static calculation, with their corresponding formation energy serving as the amorphous limit to predict synthesizability, following the method proposed by Aykol et al. \cite{aykolThermodynamicLimitSynthesis2018}.

\subsection*{Machine learning potential workflow}
The M3GNet model developed by Chen et al. \cite{chenUniversalGraphDeep2022a} offers an alternative surrogate model for AIMD computations, enabling the generation of amorphous structures and computation of Li diffusivities across various compositions. This surrogate model has been incorporated into the MPMorph workflow, serving as the calculator for energy and force. The implemented version can be accessed at \url{https://github.com/Tinaatucsd/mpmorph/blob/chgnet_fm_refactor-pv-extract/src/mpmorph/flows/md_flow.py}.

\subsection*{Random forest and XGBoost models}

For model development, we utilized random sampling to divide the multi-temperature dataset into training and test sets. The training data constitutes 85\% of the total data, with the remaining 15\% reserved for testing. The specific settings for XGBoost models are n estimators = 200, max depth = 4, n jobs = 6, and cross-validation score = negative root mean squared error. Default values are used for all other hyper-parameters of the XGBoost model and the Random Forest (RF) model implemented in the scikit-learn package. 
\subsection*{SISSO model}
A number of SISSO models\cite{ouyangSISSOCompressedsensingMethod2018} with increasing complexities were trained, and their prediction performances are shown in Figure \ref{fig: sisso models}. Some of the key input settings for SISSO training include the dimension of the descriptor desc\_dim=5; the Number of scalar features nsf=43; The parameters used to control the feature complexity rung=2 and maxcomplexity=5; The operator set considered include addition (+), subtraction (-), multiplication (*), division (/),  exponentiation such as ($^{2}$), ($^{3}$), ($^{6}$),($^{-1}$),(exp),(exp-),(log).

\section*{Acknowledgement}
This research was intellectually led by the Materials Project program (Contract No. DE-AC02-05-CH11231, KC23MP), supported by the U.S. Department of Energy, Office of Basic Energy Sciences. This study utilized the facilities of the National Energy Research Scientific Computing Center (NERSC), a User Facility of the U.S. Department of Energy Office of Science located at Lawrence Berkeley National Laboratory, operated under Contract No. DE-AC02-05-CH11231. 
\bibliography{ref}

\bibliographystyle{unsrt}

\clearpage

\end{document}



\baselineskip24pt


\maketitle

\section{Data scope}

\begin{figure}[H]
\centering\includegraphics[width=1.0\linewidth]{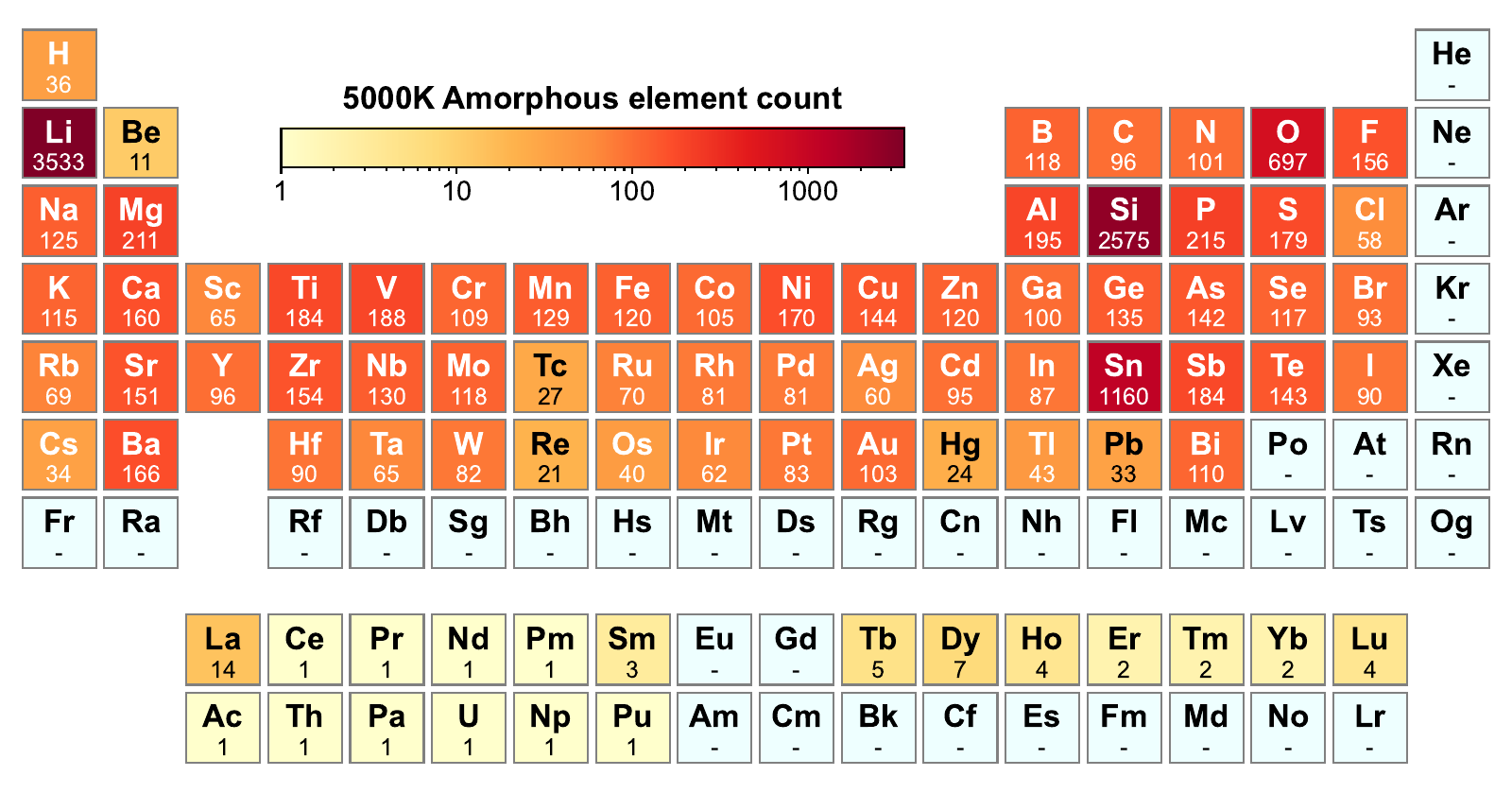}
\caption{\label{SI:high_T} Element occurrence of  compositions covered in the 5000K amorphous database}
\end{figure}

\begin{figure}[H]
\centering\includegraphics[width=1.0\linewidth]{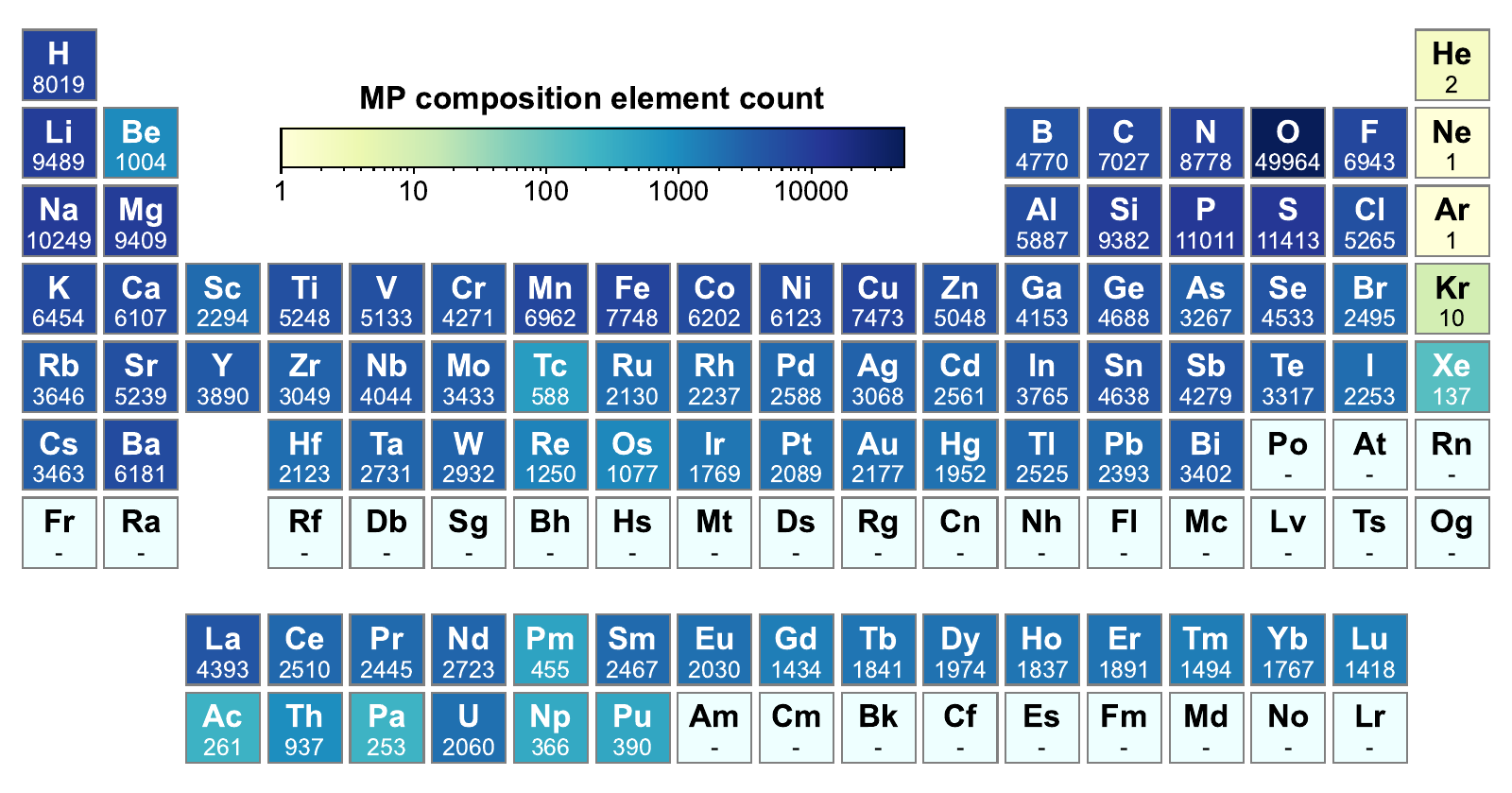}
\caption{\label{SI:MP} Element occurrence of compounds covered in Materials Project. MP entries are queried on 2023-02-07, database release 2021.05.13}
\end{figure}

\begin{figure}[H]
\centering\includegraphics[width=1.0\linewidth]{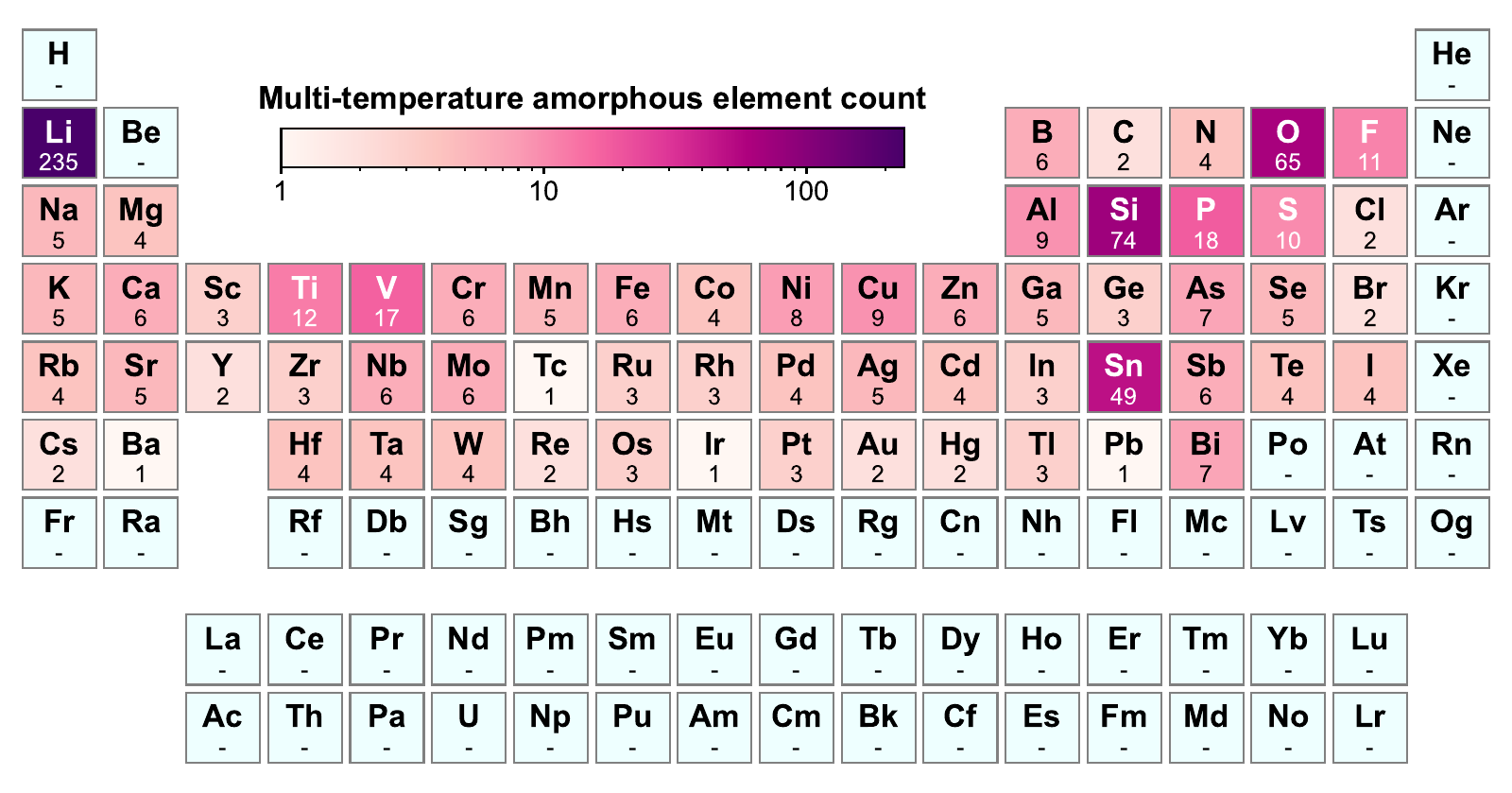}
\caption{\label{SI:multi_T} Element occurrence of  compositions covered in the multi-temperature amorphous database}
\end{figure}

\begin{figure}[htp]
\centering\includegraphics[width=1.0\linewidth]{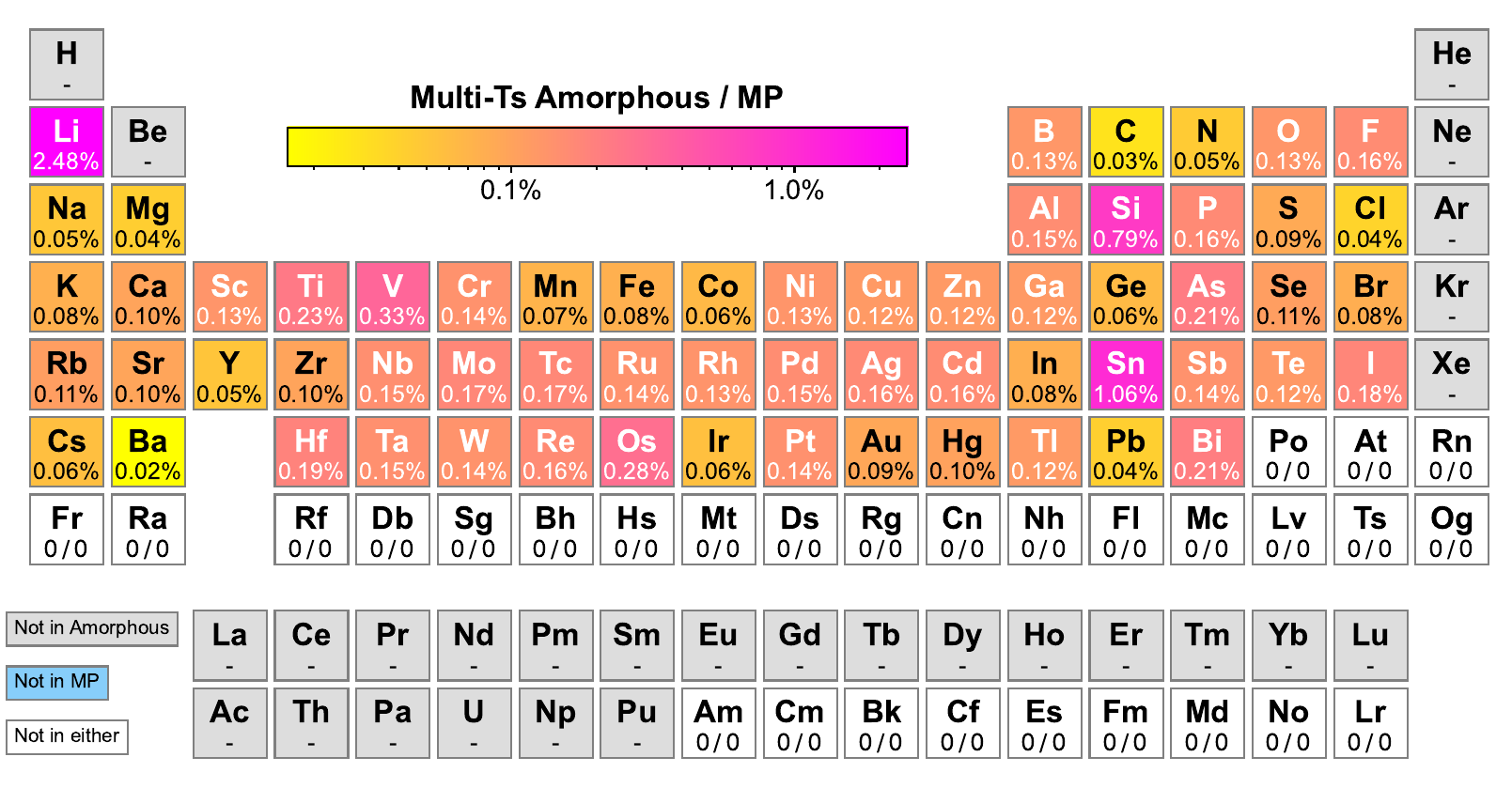}
\caption{\label{SI:multi_T_ratio} Elemental occurrence in the multi-temperature amorphous database compared to the Materials Project. Element occurrence ratios for compositions in the multi-temperature amorphous database are shaded by color.}
\end{figure}

\begin{figure}[H]
\centering\includegraphics[width=0.48\linewidth]{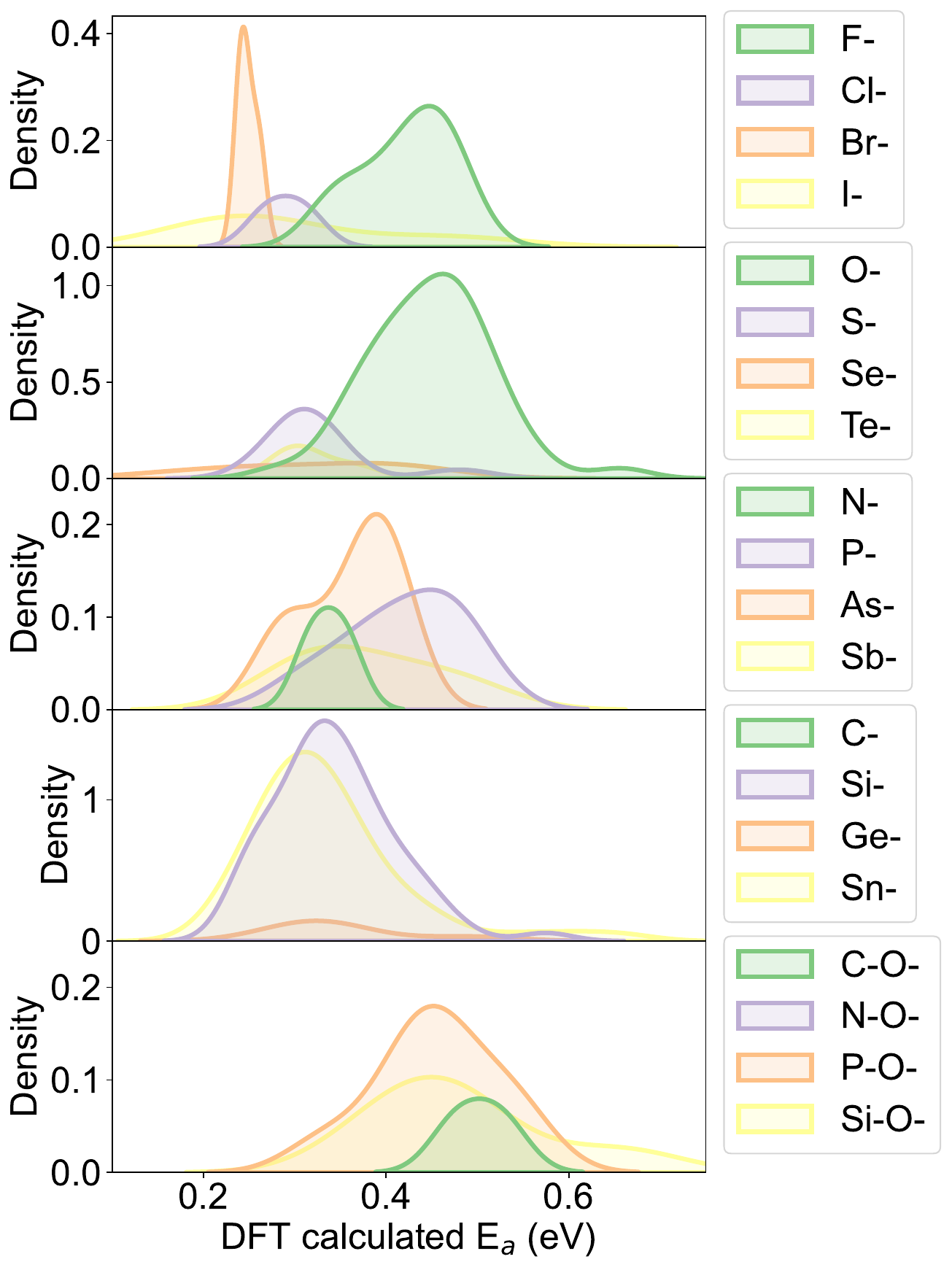}
\caption{\label{SI: Ea distribution}  \textbf{Distributions of activation energies for Li diffusion}. The kernel density estimation is used to illustrate the distribution.}
\end{figure}

\begin{figure}[H]
\centering\includegraphics[width=1.\linewidth]{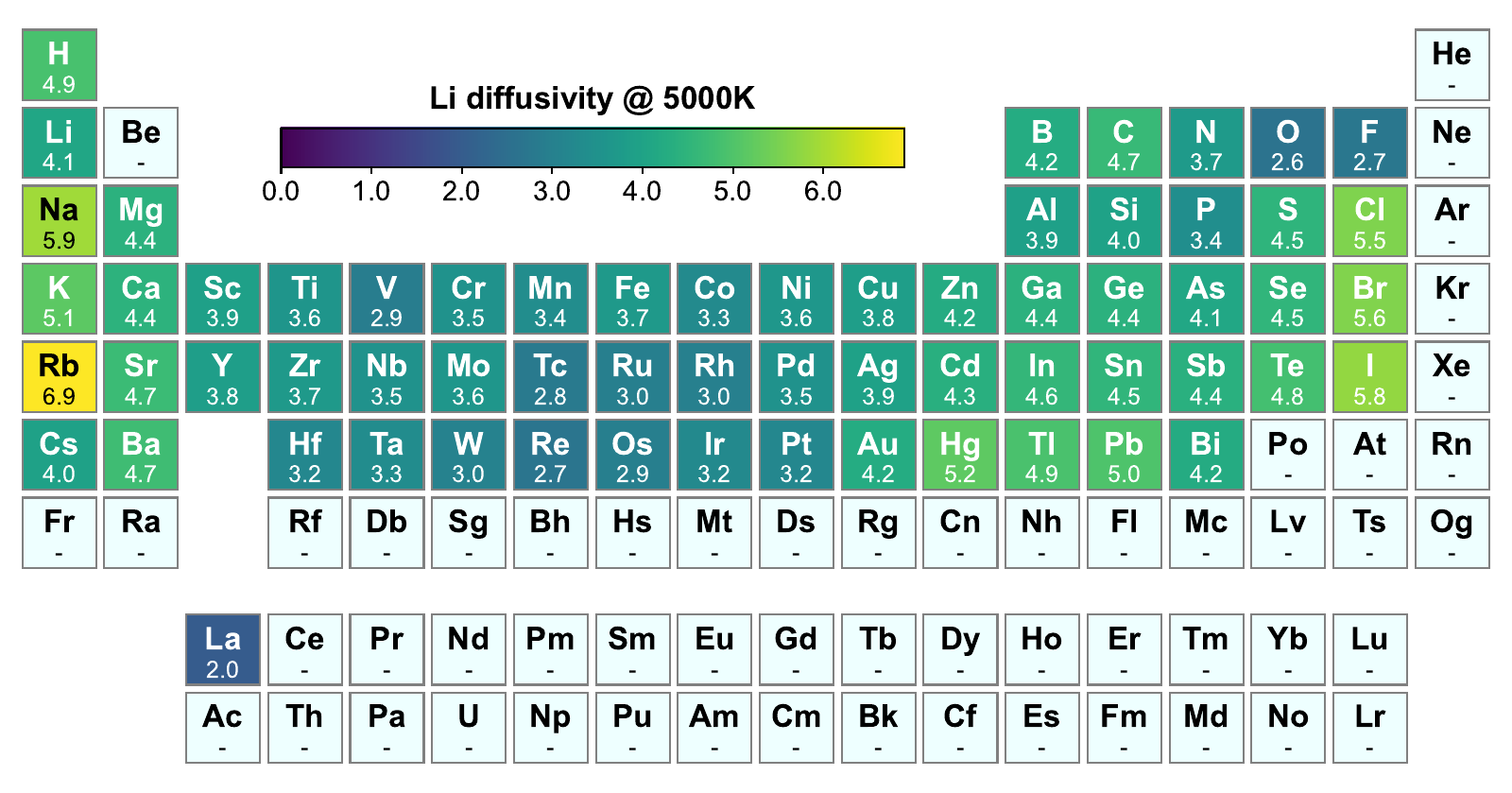}
\caption{\label{SI: high_T_mean_D} \textbf{The mean values of Li diffusivity among the compounds that include a specific element in the periodic table}. The color bar indicates that brighter hues signify greater Li diffusivity. The unit for the color bar is $\times 10^{-3} cm^2/s$}
\end{figure}

\begin{figure}[H]
\centering\includegraphics[width=1.0\linewidth]{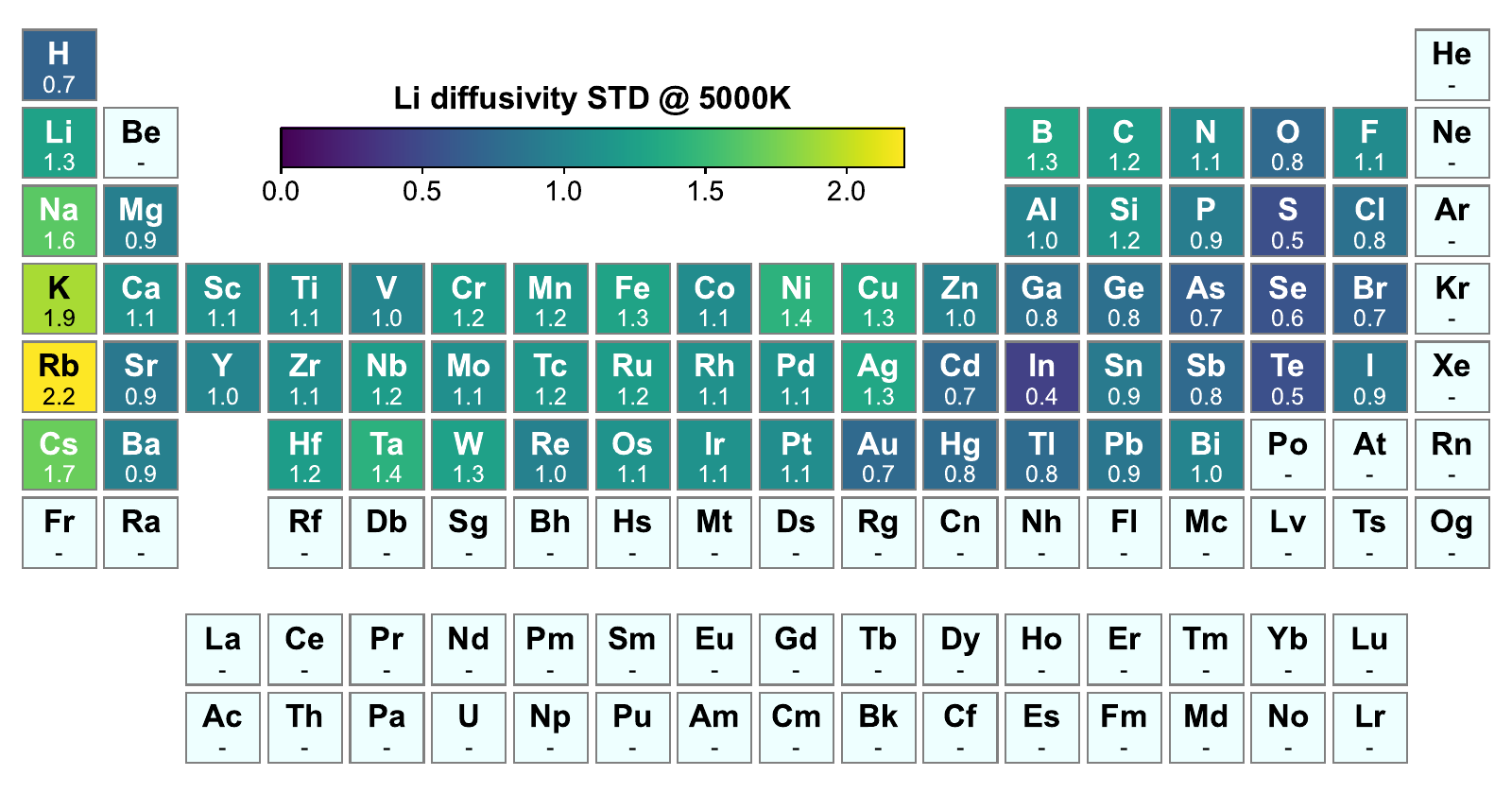}
\caption{\label{SI: high_T_std_D} \textbf{The standard deviation values of Li diffusivity among the compounds that include a specific element in the periodic table}. The unit for the color bar is $\times 10^{-3}  cm^2/s$}
\end{figure}

\begin{figure}[H]
\centering\includegraphics[width=1.0\linewidth]{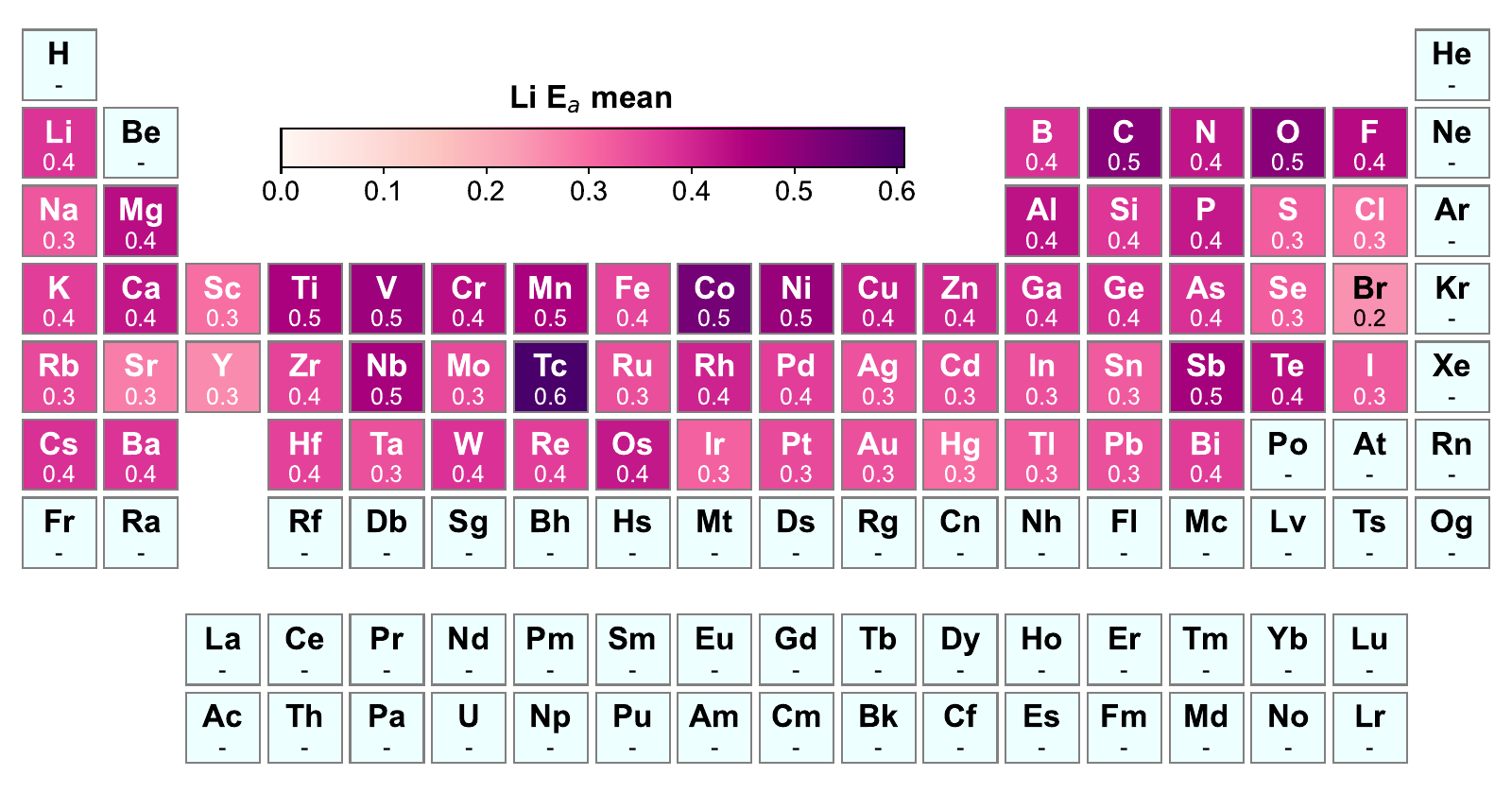}
\caption{\label{SI: Ea_mean} \textbf{The mean values of activation barrier $E_a$ for Li diffusivity among the compounds that include a specific element in the periodic table}. The unit for the color bar is eV.}
\end{figure}

\begin{figure}[H]
\centering\includegraphics[width=1.0\linewidth]{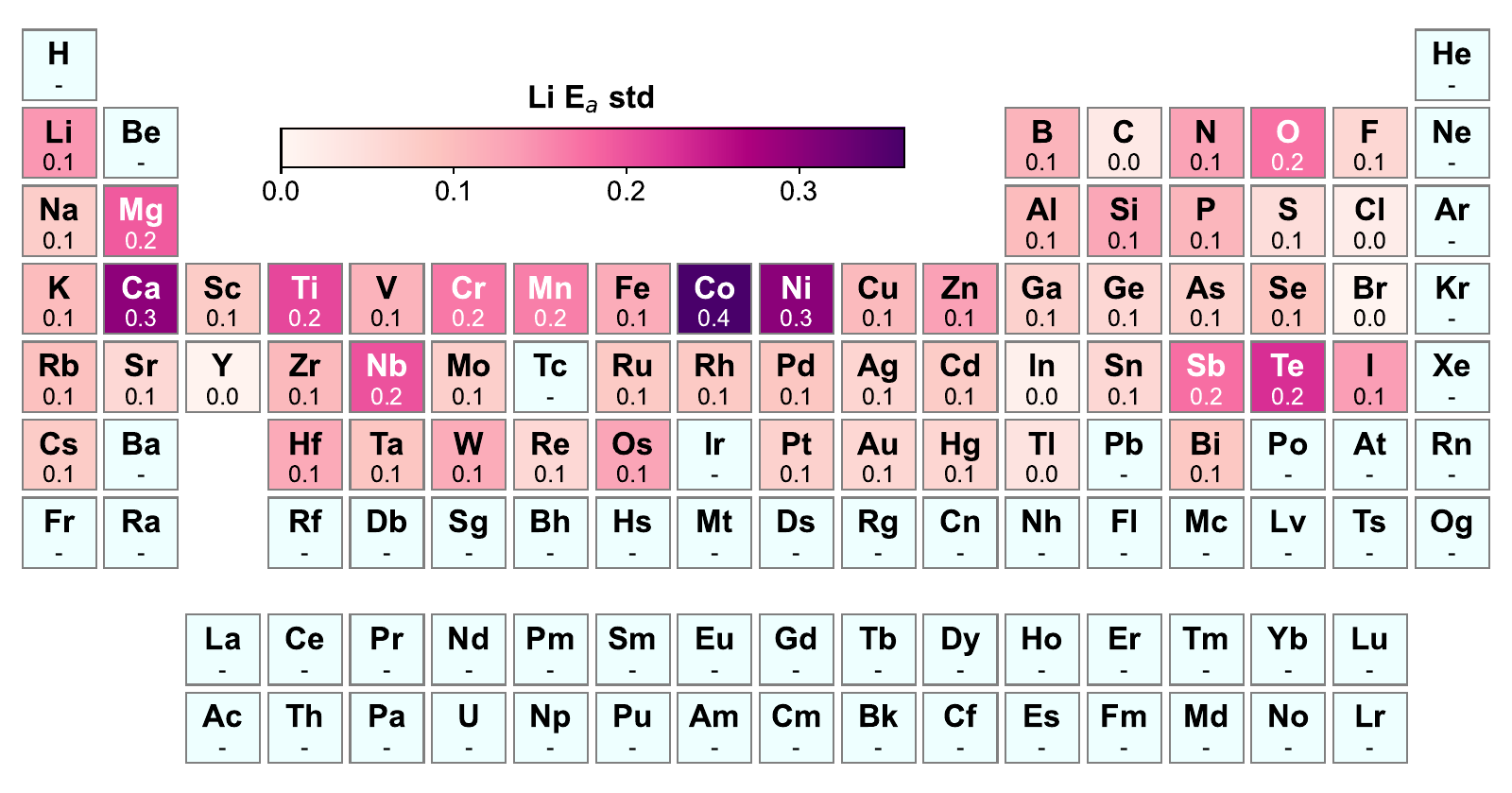}
\caption{\label{SI: Ea_std} The standard deviations of activation barrier $E_a$ for Li diffusivity among the compounds that include a specific element in the periodic table. The unit for the color bar is eV.}
\end{figure}

\begin{figure}[H]
    \centering\includegraphics[width=0.7\linewidth]{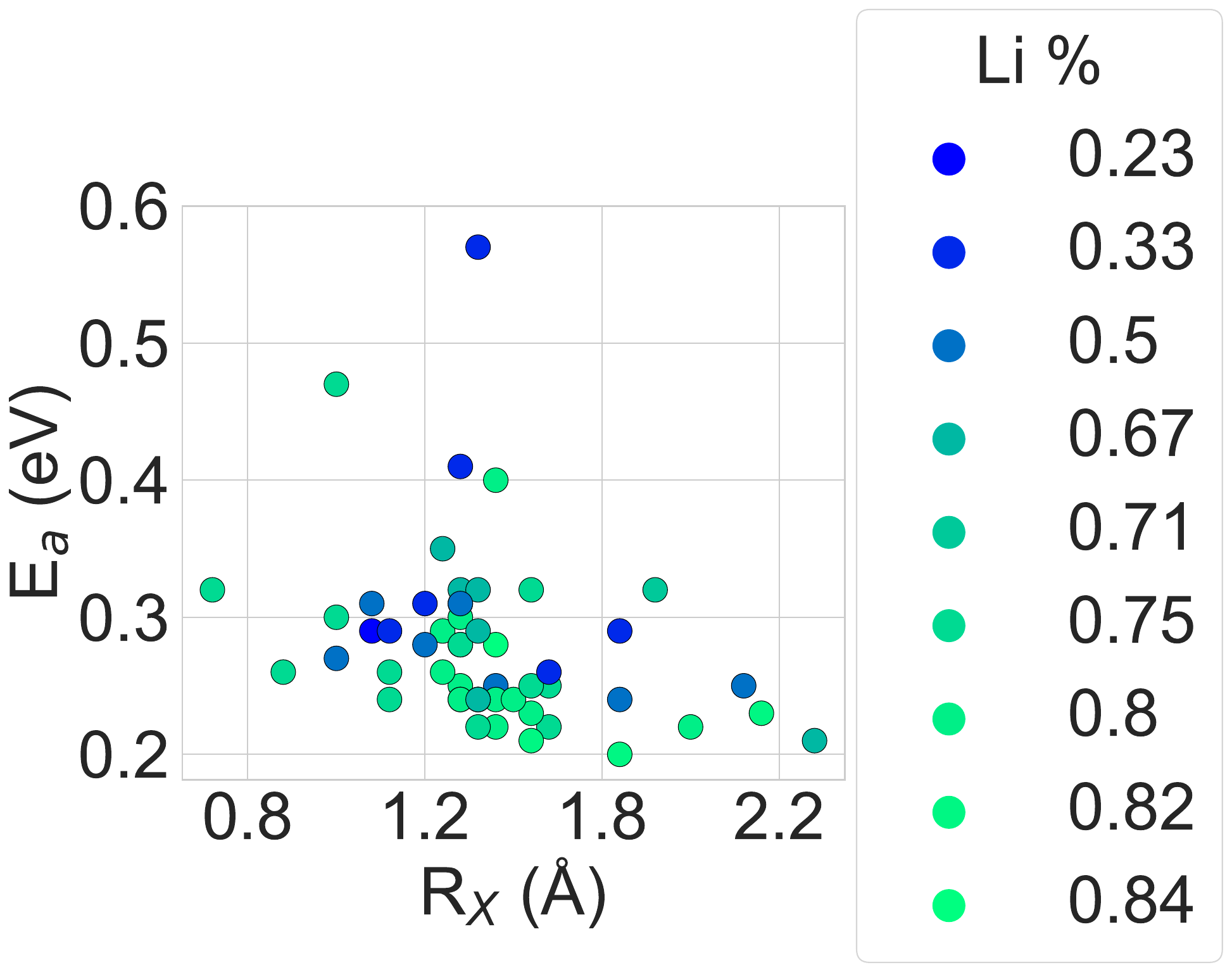}
    \caption{\label{SI: Ea_vs_Rx}  \textbf{The negative correlation between $E_a$ and $R_X$}. The color shows different Li\%, which can be used to explain why $E_a$ for certain systems are higher compared to the ones with a similar radius of element $X$.}

\end{figure}

\section{Feature calculations}

\subsection{Compositional features}

\subsubsection{The weighted average of cohesive energy}

The weighted average of cohesive energy $\overline{E_{coh}}$ and cohesive energy of composition exclude Li $\overline{E_{coh}^{Li}}$
    
The weighted average of cohesive energy ($\overline{E_{coh}}$) is determined by taking into account the cohesive energy of the constituent elemental systems in their ground states. These individual energies are then multiplied by the respective fraction of each element present in the composition. The final value is obtained by summing up these weighted energies, providing an overall representation of cohesive energy for the given composition.

    \begin{equation}
        E_{coh} = \frac{\sum_{i=1}^{i=n} E^{coh}_i\cdot w_i}{\sum_{i=1}^{i=n} w_i}
    \end{equation}

    Where $E^{coh}_i$ is the cohesive energy of the ground state of the constituent element $i$, and $w_i$ is the elemental fraction of the composition, $n$ is the number of species in the composition.

\subsection{Structural features}
 Here, signifies the. $\overline{X_{Li} - X_{others}}$ denotes the mean difference in electronegativity between Li and other elements in the compound. $\overline{R^{Li}}$ and $\overline{m^{Li}}$ respectively refer to the mean atomic radius and atomic mass of the Li neighbors. Lastly, $\frac{\Bar{R^{Li}}}{R}$ represents the ratio of the weighted average atomic radius of the Li neighbors to the weighted average atomic radius of the compound.

\begin{figure}[H]
\centering\includegraphics[width=1.0\linewidth]{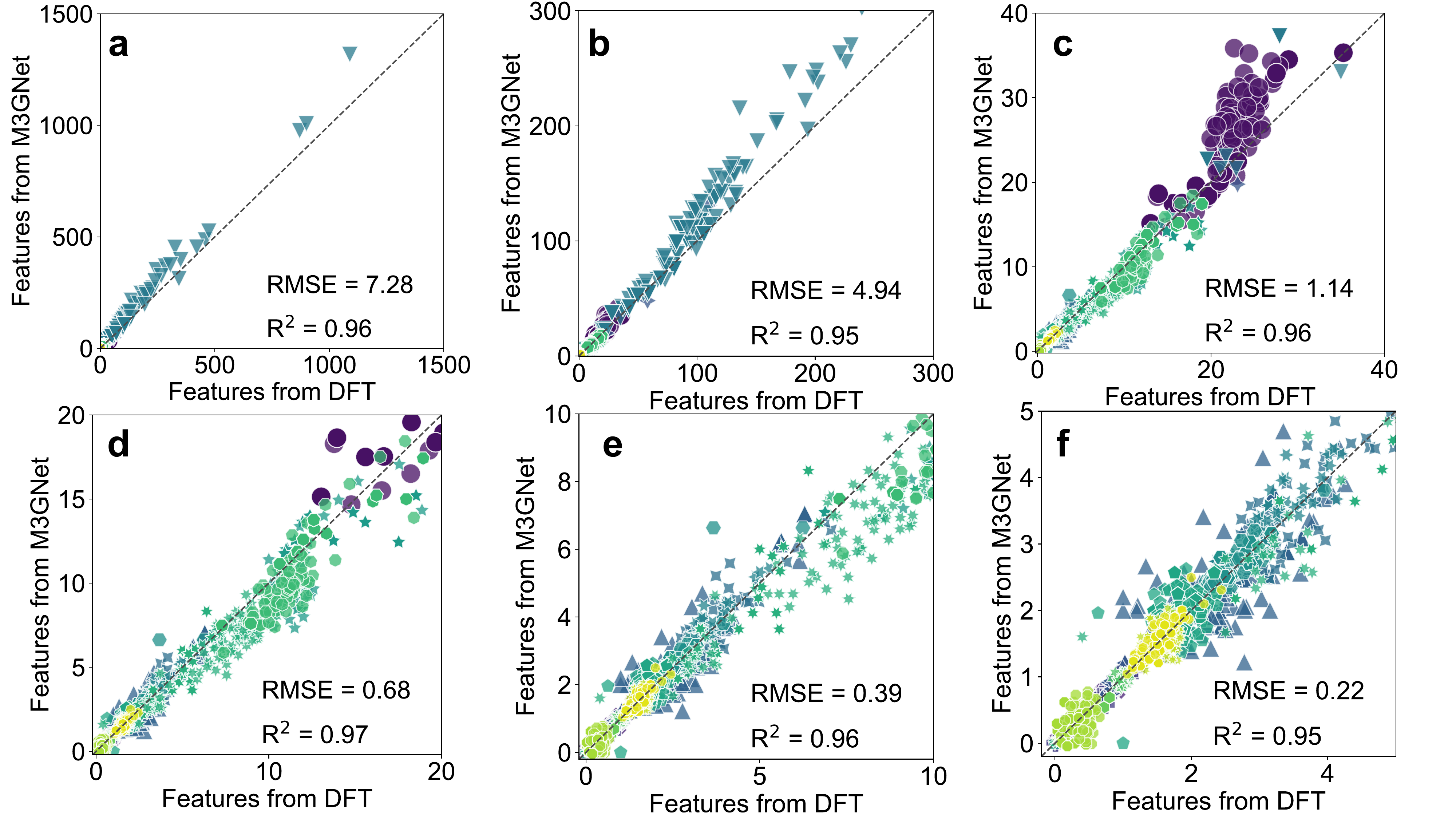}
\caption{\label{SI: structure feature}  The parity plots of structure features calculated from AIMD and M3GNET.}
\end{figure}


\begin{figure}[H]
\centering\includegraphics[width=1.0\linewidth]{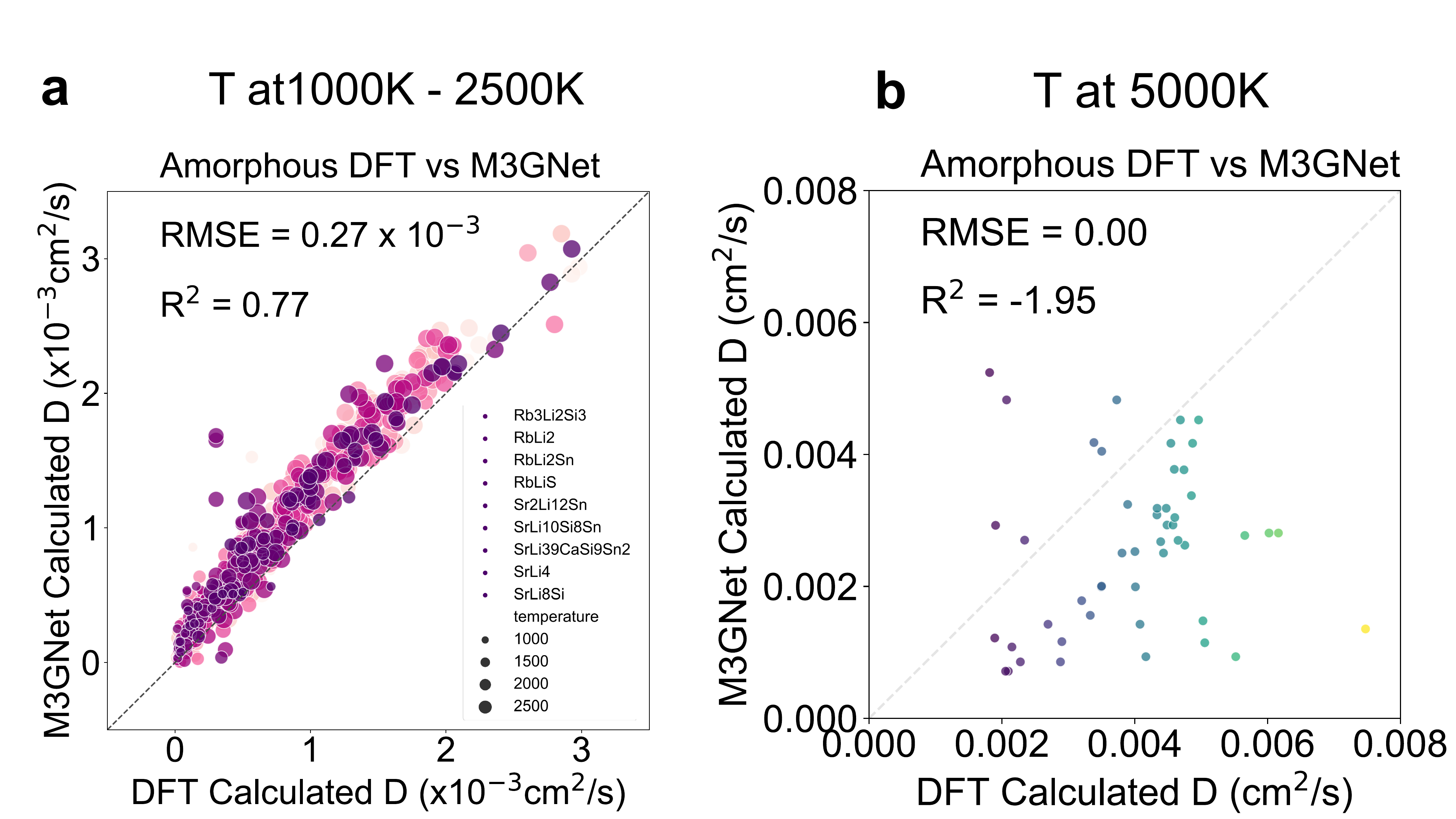}
\caption{\label{SI: diffusivity parity} The parity plot of the Li diffusivity calculated from AIMD and M3GNET at intermediate temperatures (1000K, 1500K, 2000K, and 2500K) and ultra-high temperature at 5000K.}
\end{figure}